\documentclass[pdflatex,sn-basic]{sn-jnl}%

\usepackage{graphicx}
\usepackage{multirow}
\usepackage{amsmath,amssymb,amsfonts}
\usepackage{amsthm}
\usepackage[title]{appendix}
\usepackage{xcolor}
\usepackage{textcomp}
\usepackage{manyfoot}
\usepackage{booktabs}
\usepackage{algorithm}
\usepackage{algorithmicx}
\usepackage{algpseudocode}
\usepackage{listings}
\usepackage{subcaption}
\usepackage{hyperref}
\usepackage{longtable, geometry}
\usepackage{tabularx,ragged2e}
\usepackage{booktabs}
\usepackage[T1]{fontenc}
\usepackage[english]{babel}
\usepackage[font=small,labelfont=bf,labelsep=space]{caption}
\usepackage{tablefootnote}
\usepackage{enumitem}

\newcommand{\popScoreCi}{\rho(c_{i})}
\newcommand{\subscript}[2]{$#1 _ #2$}
\newcommand{\CoTwo}{CO\textsubscript{2}}
\newcommand{\COtwoE}{\text{CO\textsubscript{2}e}}
\newcommand{\sigmaCij}{\sigma(c_{i}^{j})}
\newcommand{\EmissionScorei}{\mathit{Z(c_{i})}}
\newcommand{\Cityi}{\mathit{c_{i}}}

\newcommand{\alphaCost}{\alpha_{Cost}}
\newcommand{\betaPOI}{\beta_{POI}}

\newcommand{\SFindexCij}{SFI(c_{i}^{j})}
\newcommand{\SF}{\textit{S-Fairness}}
\newcommand{\SFI}{\textit{S-Fairness Indicator}}
\newcommand{\minHighlight}[1]{\underline{\textit{#1}}}
\newcommand{\maxHighlight}[1]{\textbf{#1}}

\newcommand{\rebuttal}[1]{\textcolor{black}{#1}}
\newcommand{\tunar}[1]{\textcolor{black}{#1}}

\newcommand{\theadb}[2]{%
  \multicolumn{#1}{c}{\bfseries\begin{tabular}[t]{@{}c@{}} #2\end{tabular}}%
}

\addto\extrasenglish{%
}
\begin{document}

\title[Modeling Sustainable City Trips]{Modeling Sustainable City Trips: Integrating \tunar{\COtwoE} Emissions, Popularity, and Seasonality into Tourism Recommender Systems}

\author{\fnm{Ashmi} \sur{Banerjee}\email{ashmi.banerjee@tum.de}}
\author{\fnm{Tunar} \sur{Mahmudov}\email{tunar.mahmudov@tum.de}}
\author{\fnm{Emil} \sur{Adler}\email{emil.adler@tum.de}}
\author{\fnm{Fitri} \sur{Nur Aisyah}\email{fitri.aisyah@tum.de}}
\author{\fnm{Wolfgang} \sur{W\"orndl}\email{woerndl@in.tum.de}}

\affil{\orgdiv{TUM School of Computation, Information and Technology}, \orgname{Technical University of Munich}, \orgaddress{\street{Boltzmannstrasse 3}, \city{Garching}, \postcode{85748}, \country{Germany}}}

\abstract{
\rebuttal{
Tourism affects not only the tourism industry but also society and stakeholders such as the environment, local businesses, and residents as well.
Tourism Recommender Systems (TRS) can be pivotal in promoting sustainable tourism by guiding travelers toward destinations with minimal negative impact.
Our paper introduces a composite sustainability indicator for a city trip TRS based on the users’ starting point and month of travel.
This indicator integrates CO2e emissions for different transportation modes and analyses destination popularity and seasonal demand.
We quantify city popularity based on user reviews, points of interest, and search trends from Tripadvisor and Google Trends data. 
To calculate a seasonal demand index, we leverage data from TourMIS and Airbnb. 
We conducted a user study to explore the fundamental trade-offs in travel decision-making and determine the weights for our proposed indicator.
Finally, we demonstrate the integration of this indicator into a TRS, illustrating its ability to deliver sustainable city trip recommendations.
This work lays the foundation for future research by integrating sustainability measures and contributing to responsible recommendations by TRS.}}


\maketitle
\section{Introduction} \label{section: introduction}

Recommender Systems (RS) provide tailored content to individual preferences, spanning diverse domains like e-commerce, social media, news, and more, effectively managing information to prevent overload~\citep{abdollahpouri2020multistakeholder}. In travel and tourism, RS is pivotal in simplifying trip planning by providing personalized recommendations for destinations, accommodations, activities, and more~\citep{ISINKAYE2015261}. However, this is a particularly challenging domain due to the influence of dynamic factors such as seasonality and travel regulations~\citep{balakrishnan2021multistakeholder}, as well as constraints related to capacity-limited resources such as airline seats, hotel rooms, and event tickets~\citep{abdollahpouri2021multistakeholder}.

Traditionally, RS focused on delivering accurate user recommendations, but in practice, they function as a convergence point for multiple stakeholders, making it a multistakeholder scenario~\citep{abdollahpouri2020multistakeholder}. Recognizing the interests of all stakeholders becomes crucial in this dynamic.
Our stakeholder classification, inspired by~\citet{balakrishnan2021multistakeholder}, identifies four key categories: consumers, item providers, platform, and society, aligning with common touristic recommendation scenarios. 
Despite this seemingly straightforward categorization, real-world stakeholder relationships are often more intricate. 
Each stakeholder is vested in the traveler's journey, and optimizing consumer recommendations can yield benefits for all involved parties~\citep{abdollahpouri2020multistakeholder}.
Complexities arise when the goals of stakeholders conflict, \rebuttal{mainly as profit motives often drive them, leading to inevitable trade-offs in achieving fairness in these systems~\citep{Jannach}.
This necessitates adopting a multistakeholder approach, acknowledging stakeholders' interdependence, and balancing their objectives when designing fair Tourism Recommender Systems (TRS).}

Tourism's impact goes beyond active participants; it affects the local environment and businesses and profoundly influences the balance of nature. Thus, developing a fair TRS involves recommending sustainable options and fostering responsible tourism practices.  
World Tourism Organization and United Nations Development Programme define \textit{sustainable tourism} as "\textit{tourism that takes full account of its current and future economic, social and environmental impacts, addressing the needs of visitors, the industry, the environment, and host communities}" ~\citep{gossling2017tourism}.
Achieving sustainability in tourism requires interventions at various levels, including municipal policies and regulations~\citep{werthner2015future}.
However, measuring sustainability at destinations poses a significant challenge, impeding effective decision-making, management, and meeting destination needs~\citep{fernandez2009measuring}.
In the context of tourism, a destination is characterized by \textit{"a country, state, region, city, or town that is actively marketed or markets itself as an appealing place for tourists to visit"}~\citep{beirman2020restoring}. 
A destination's sustainability is crucial for long-term competitiveness and visitor satisfaction, and should not solely be determined by arrival numbers or bed nights~\citep{onder2017tourmis}.

\rebuttal{One of the essential interventions where a well-designed TRS can play a vital role is regulating the number of tourists.
A TRS can be especially helpful in addressing the challenges of over- and undertourism, both of which are on the rise due to factors such as low-cost aviation, affordable transportation, social media influence, and platforms like Airbnb~\footnote{https://www.airbnb.com/}~\citep{Gowreesunkar2020}.}
Overtourism, witnessed in popular destinations like Venice, Barcelona, Rome, and Dubrovnik, poses threats to historic preservation, the environment, residents, and overall tourist experiences, making it challenging to find reasonably priced housing in these cities~\citep{dastgerdi2023post,dodds2019phenomena}. Conversely, undertourism, prevalent in under-explored destinations, results from a lack of infrastructure, publicity, and accessibility. Both scenarios have adverse consequences.
For instance, the recent COVID-19 pandemic highlighted the adverse effects of undertourism, causing significant disruptions to the tourism and hotel industries~\citep{gali2022impacts}. To address these issues, a TRS must be designed to provide responsible recommendations, considering the interests of all stakeholders. These systems should advocate for sustainable tourism practices, promoting responsible tourism while offering personalized suggestions to users.
This involves recommendations encouraging tourists to visit destinations with minimal environmental impact, promoting less popular yet attractive locations, and balancing the tourist load uniformly throughout the year.

A substantial amount of research has been conducted on developing fair recommendation systems that consider the interests of all stakeholders involved in tourism ~\citep{rahmaniPOI,shen2021sar,weydemann2019defining,wu2021tfrom}. 
However, there has been limited focus on generating sustainable recommendations~\citep{ashmi2023survey}. 
This paper explores the concept of modeling \textit{Societal Fairness} or \tunar{\SF}. It emphasizes the impact of tourism on individuals who are not directly involved, such as residents, environment, and other stakeholders, collectively referred to as \textit{"society"}~\citep{ashmi_umap_dc_2023}. These stakeholders often encounter challenges such as rising housing prices, environmental pollution, and traffic congestion due to heightened tourism activities in their vicinity.

\rebuttal{In this paper, we aim to help travelers seeking recommendations for their vacations in European cities based on their starting points. We compiled a list of the 200 most densely populated European cities to be considered destinations. Our proposed method aims to assess the sustainability of these destinations based on the month of travel and the user's starting point. To achieve this, we assign a composite \tunar{\SFI} to all cities that are accessible from the user's initial location.}
To mitigate the adverse effects on the environment and society, we identify three key factors in the calculation of the \tunar{\SFI}:

    \begin{enumerate}
        \item Destinations with environmentally friendly travel options, minimizing~\tunar{\COtwoE}~emissions incurred during the travel to the destination.
        \item Suggesting less popular yet attractive destinations.
        \item Choosing destinations with lower demand during the specific travel month.
    \end{enumerate}

Our approach is designed to assist travelers in making more sustainable travel choices.  
\rebuttal{The \tunar{\SFI} aggregates individual assessments of \tunar{\COtwoE}~emissions, popularity and monthly seasonality. 
A lower number represents less impact on local communities and the environment.
It is important to note that while there are existing methods to measure the individual components of the \tunar{\SFI}, our novelty lies in integrating them into a single metric to measure sustainability or \tunar{\SF}. This metric can then be incorporated into TRS to recommend sustainable destinations to users.}

To this end, our work makes the following contributions:

\begin{enumerate}
    \item Defining the concept of sustainability and its elements within a city trip recommender system.
    \item Collecting and examining data on transportation~\tunar{\COtwoE}~emissions, as well as determining popularity and seasonality indices for cities. 
    \item \rebuttal{Assigning a composite \tunar{\SFI} to cities based on a specific starting point and month of travel.}
    \item Executing a user study to explore fundamental travel trade-offs and ascertain the \tunar{\SFI} weights.
    \item \rebuttal{Demonstrating how our proposed indicator can be integrated into a TRS to provide sustainable city trip recommendations to users.}
\end{enumerate}

Our paper is structured as follows:\rebuttal{~\autoref{section: related} reviews prior research on city trip recommendations and sustainability in TRS. 
~\autoref{section: transportation},~\autoref{section: popularity}, and~\autoref{section: seasonality}, present detailed methodologies for calculating the individual components of the composite \tunar{\SFI} score --- transportation-related~\tunar{\COtwoE}~emissions, city popularity, and seasonal demand, respectively.
In~\autoref{section: user_study}, we outline the findings from our user study on factors influencing destination choices and openness to sustainable tourism recommendations.}
~\autoref{section: s-fairness} elucidates the concept of \textit{Societal Fairness}, assigns the relevant indices, and \rebuttal{subsequently describes how our indicator can be integrated into TRS through a dedicated user study.}
Finally, \autoref{section: conclusion} concludes the paper, summarizing key findings and suggesting future studies in this area.

\section{Related Work} \label{section: related}

\rebuttal{In the context of this study, we explore the related work along two dimensions -- \textit{City Trip Recommendations} in~\autoref{subsection: rw_city_trip_recos} and \textit{Sustainability in Tourism Recommender Systems} in~\autoref{subsection: rw_sustainable_tourism}.
This reflects the shift in recommender systems towards integrating personalization and user preferences in city trip recommendations while considering sustainability and societal fairness.}

\subsection{City Trip Recommendations} \label{subsection: rw_city_trip_recos}
\rebuttal{City trip recommender systems are essential not only to simplify trip planning for users but also to cater to the evolving needs and challenges in the travel domain. 
Traditionally, recommender systems have predominantly centered on modeling user preferences to offer personalized suggestions. However, this domain presents distinctive challenges, including the intangibility of recommended items and dynamic factors such as seasonality, travel regulations, and resource constraints~\citep{balakrishnan2021multistakeholder, werthner2004commerce}. Recent research has emphasized addressing these aspects as well.}

\rebuttal{Besides techniques like collaborative filtering leveraging users' past activities, similarities with other users, and network-based preferences are also prevalent in literature~\citep{Lu2012, Dadoun2019,Pu2020}.
Constraints defined by users, such as budget or time preferences, have also been addressed as critical components in tailoring travel packages~\citep{xie2010breaking,lim2018personalized}. However, challenges like the cold start problem and data sparsity have led to the adoption of content-based approaches. These approaches construct domain models using relevant features for tourism, often derived from expert opinions, literature, or data-driven methods~\citep{liu2011personalized,Pu2020}.}

\rebuttal{While expert-driven models offer nuanced insights, their cost and complexity necessitate complementing them with diverse data sources, such as Location-based Social Networks (LBSNs) for venues~\citep{Lu2012, Dadoun2019}.
For instance, the data-driven system introduced by~\citet{roy2021triprec} tailors composite city trips to users based on their preferences and constraints using data from the LSBNs.
It suggests personalized itineraries for users by gathering information such as their home region, travel duration, and venue preferences.
Similarly, the work by~\cite{myftija2020cityrec} uses a data-driven characterization of cities inside a conversational recommender system to recommend cities as destinations.
~\cite{dietz2018recommending} present wOndary, a platform designed for global trip planning and sharing via crowdsourcing. Using content-based recommendation methods and a structured itinerary representation, wOndary tackles issues surrounding item discovery and routing in tourist trip design.
~\citet{massimo2019clustering}, explores the concept of clustering users with similar POI visit trajectories and then constructing a general user behavior model via inverse reinforcement learning. This model allows for generating recommendations based on learned behavioral patterns.}

\subsection{Sustainability in Tourism Recommender Systems} \label{subsection: rw_sustainable_tourism}

\textbf{Measuring Sustainability:}
\rebuttal{Sustainable tourism presents complexities in defining objectives and indicators for a sustainable TRS.~\citet{ko2005development} assert that sustainability concerns vary by destination, necessitating specific dimensions and methodologies tailored to each location. However, this tailored approach raises comparability issues, as different sets of indicators hinder meaningful comparisons between destinations. In response, ~\citet{cernat2012paths} develop the Sustainable Tourism Benchmarking Tool (STBT), comprising 54 indicators across seven dimensions. Though tested in 75 countries, the tool performs well only in Indonesia, Malaysia, and Thailand, where data availability is higher. 
\citet{onder2017tourmis} suggest prioritizing the analysis of existing sustainable tourism indicators over introducing new, less practically applicable measures. They propose utilizing data envelopment analysis (DEA) on existing tourism information systems to model destination competitiveness and create a utility function sustainability index. DEA enables the identification of a destination's efficiency and facilitates benchmarking against others to pinpoint areas for improvement. While this index is valuable for guiding policymakers and city developers in enhancing specific aspects of cities, it has not been integrated into TRS yet.}

\rebuttal{~\citet{gorantla2023towards} utilize the Circles of Sustainability framework to assess a city's sustainability across four domains and seven subdomains each, aggregating data sources to compute a simplified sustainability index, providing valuable insights into city sustainability in comparison to others.
In contrast,~\citet{hoffmann2022measuring} adopt a fully data-driven approach, analyzing Tripadvisor data to compare sustainability measures among hotels. Their unsupervised statistical learning approach yields improved performance in classifying sustainable hotels. However, the model's explainability is limited; while it indicates larger hotels are more likely to be sustainable, causality cannot be explicitly inferred.}

\textbf{Sustainable Recommendations:} \rebuttal{While numerous studies explore the quantification of sustainability in tourism, there is a notable scarcity of their application within recommender systems in this domain.
Sustainable recommender system seeks to create a balance across economic, environmental, and social dimensions which is essential for creating a resilient ecosystem~\citep{gorantla2023towards}. One can initiate the integration of features that capture user interest while simultaneously promoting sustainability.}
\rebuttal{However, a sustainable recommender system should not be fully user-centric. The tourism industry involves different stakeholders, including consumers, providers, platforms, and society, thus it is natural for sustainable RS to implement a multistakeholder approach. In a recent examination of multistakeholder fairness within tourism recommender systems, it became apparent that existing studies predominantly concentrate on provider and consumer stakeholders. Surprisingly, the societal aspect, despite bearing significant impacts from tourism, is often overlooked as a stakeholder~\citep{ashmi2023survey}.}

\rebuttal{A recent study by~\citet{merinov2022sustainability} introduces a multistakeholder model that not only considers user preferences but also the occupancy level of the destination. In another study by~\citet{patro2020towards}, the multi-objective model focuses on providers' sustainability by maintaining their exposure while also preventing overcrowding.~\citet{pachot2021multiobjective} added local authorities as stakeholders, with economic growth, productive resilience, prioritizing basic necessities, and greener production as its objective. 
While~\citet{banik2023fairness} address overcrowding by promoting fair recommendations that encourage users to choose greener alternatives, they lack a specific method for measuring the generation of recommended items.}

\rebuttal{Our methodology differs from the state-of-the-art in terms of data collection and analysis techniques. 
We employ a combination of qualitative insights from participant feedback and quantitative data from various sources to assign a sustainability metric to the destinations based on the users' origin.
This approach allows for a more nuanced understanding of tourist behavior and preferences, leading to more effective recommendations for city trips.}

\section{Destinations, Transportation, and Emission Estimations}\label{section: transportation}

\rebuttal{Tourism contributes to approximately 8\% of global \tunar{\COtwoE} emissions, stemming from various sources such as accommodation, food consumption, shopping, services, and agriculture. 
However, research has shown that transportation modes to and from destinations are the most significant contributors to the \tunar{\COtwoE} footprint, making up 49\% of emissions. This highlights the importance of considering the emissions from transportation as one of the key sustainability indicators in tourism, especially for city policymakers~\citep{lenzen2018carbon}.
These transportation-related emissions often result from a combination of modes chosen by travelers, the distance traveled, and the average length of stay~\citep{onder2017tourmis}.}
For example, longer-distance trips that necessitate air travel between cities tend to incur higher emissions, posing a more significant environmental impact than shorter trips feasible with public transport like trains. Conversely, a short-distance trip by car may result in more emissions than a train.

This paper adopts the estimation of greenhouse gases (GHG) emitted by various transportation modes to indicate the trip's environmental responsibility concerning the travelers' starting point. 
\rebuttal{In~\autoref{section: extracting cities}, we elaborate on the methodology employed for selecting our destinations, while~\autoref{section: data gathering ghg} outlines the detailed process of gathering data for various transportation modes.
We examine the trade-off between travel times, transportation modes, their respective emissions, and the associated costs in~\autoref{section: ghg estimations} by assigning an \tunar{\textit{emissions trade-off index}} $\EmissionScorei$ to each city $\Cityi$ reachable from the user's point of origin.}
A lower $\EmissionScorei$ indicates a more environmentally friendly and responsible tourism approach.

Our approach encourages travelers to opt for public transport, particularly for shorter to medium-distance destinations. 
Analyzing trade-off values provides insights into user behavior, enabling the formulation of effective policies. These policies might include strategies to improve travel time on routes with lower emissions but higher costs. 

Subsequent sections provide detailed discussions on these aspects.

\subsection{Extracting Destinations}\label{section: extracting cities}

\begin{figure}[htbp]
    \centering
    \includegraphics[width=\textwidth]{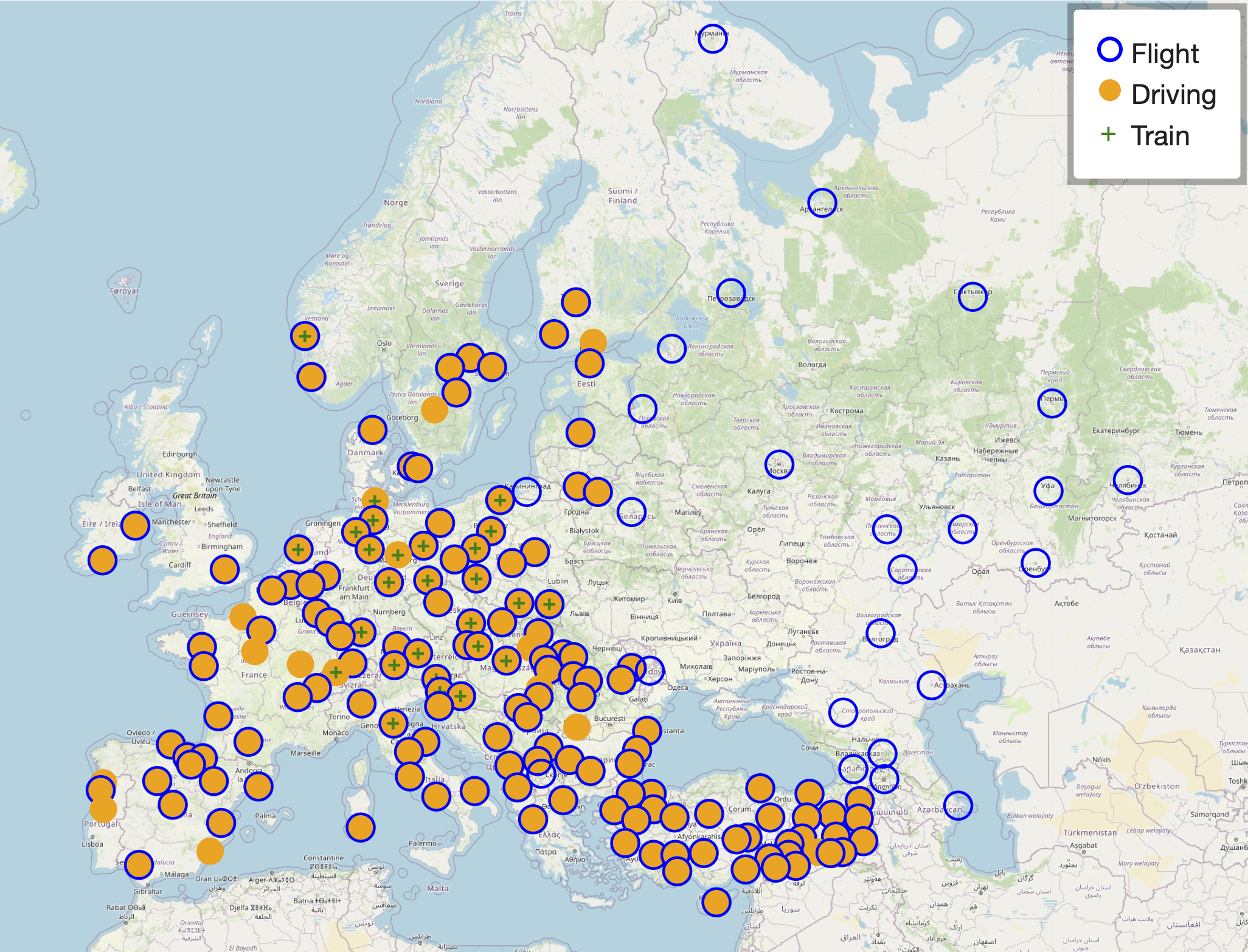}
    \caption{\rebuttal{The geographical distribution of 200 European cities with at least one airport is depicted in blue, cities prioritized for driving are highlighted in orange, and those considered for their train networks are shown with a green +} 
    }
    \label{fig:european-cities}
  \end{figure}

\rebuttal{In our scenario, a traveler is seeking a suitable city to visit for vacation from a specified starting location. They are presented with a list of European cities as potential destinations.}
The initial step involves collecting data on potential destinations for city trips.
\rebuttal{Our item space comprises 200 European cities or destinations spanning 43 countries.
We have chosen these destinations due to the continent's extensive connectivity via various modes of transportation, including flights, rail, and road, making it a highly popular destination among tourists.
Data for European cities is sourced from the world cities database~\citep{simplemaps}, filtering for the top 200 most populated cities, each featuring at least one airport.}
Subsequently, we calculate transportation emissions for travel to each city using three modes --- flights, driving, and rail where applicable, as explained below. 
\autoref{fig:european-cities} illustrates the geographical dispersion of the 200 European cities considered for our analysis. It also shows the subset of cities specifically chosen for in-depth examination regarding driving and train connections in this study.

\subsection{Data Gathering: Transportation} \label{section: data gathering ghg}

\rebuttal{This section outlines the detailed data gathering process for each transportation mode across all the cities identified in~\autoref{section: extracting cities}.}

\subsubsection{Flights}
In obtaining flight information, a major step involves associating cities with their respective airport details and IATA~\footnote{https://en.wikipedia.org/wiki/IATA\_airport\_code} codes. 
We gathered data from the Flugzeuginfo.net~\footnote{https://www.flugzeuginfo.net/} website, aligning the information with our city list. 
Notably, cities featuring multiple international airports were also incorporated into the dataset. 
This resulted in a compilation of 222 unique airports across 200 cities.

We established connections between airports $\mathit{a_{1}}$ and $\mathit{a_{2}}$, corresponding to cities $\mathit{c_{1}}$ and $\mathit{c_{2}}$, respectively. 
A dedicated connection was established for each unique route between the two cities, resulting in 16,261 unique one-way routes. 
Subsequently, each connection served as input for querying Google Flights~\footnote{https://www.google.com/travel/flights}, allowing us to extract detailed information. We prioritized the best departing flight options based on criteria such as travel time and number of stops for each trip, focusing on economy class. 

We classified each route based on the Eurocontrol's~\footnote{https://www.eurocontrol.int/} definition for distances, segmenting them into four categories: very short haul, short haul, medium haul, and long haul. Specifically, very short-haul flights covered less than 500 km, short-haul flights ranged from 500 to 1,500 km, medium-haul flights spanned 1,500 to 4,000 km, and long-haul flights exceeded 4,000 km in flying distance.
The scraped data from Google Flights lacked information on the distances between the two cities. Therefore, we employed the Great Circle Distance (GCD)~\citep{greatcircledistance, haversine_formula} measurement to calculate the distances for each route.
As indicated in~\autoref{tab:transportation-summary}, our dataset predominantly comprised short-haul and medium-haul flights, with a comparatively smaller number of long-haul flights.

\subsubsection{Driving}

We assume that driving between cities is not viable when the distance exceeds 1,000 kilometers, so we eliminate the longer connections between city pairs.
This refinement yielded 10,056 unique (two-way) connections spanning 200 European cities.
Using two prominent data sources, we derived these connections' driving distance and travel time. 

Initially, we leveraged the Google Maps Routes API~\footnote{https://developers.google.com/maps/documentation/routes} in the eco-friendly mode to acquire the most fuel-efficient driving distance and time, factoring in real-time traffic conditions, obtaining the details for 4,718 unique one-way routes. 
Subsequently, we utilized the Open Source Routing Machine (OSRM) API~\footnote{https://project-osrm.org} in driving mode to compute the driving distance and time between connections. OSRM, an open-source routing engine, relies on OpenStreetMap~\footnote{https://www.openstreetmap.org} data, providing information for all 1056 connections in our dataset. The overview of both datasets can be found in~\autoref{tab:transportation-summary}.

Both datasets contained data regarding distance, travel time, and specific route details between two cities. Furthermore, the Google Maps Routes API offered an estimate of the fuel consumption for each route. Upon conducting a comparative analysis, it was observed that Google and OSRM demonstrated a mean absolute percentage difference of 5.63\% in their respective distance calculations.
This divergence may be attributed to the different methodologies employed by Google and OSRM in estimating distances, considering factors such as routes and traffic conditions. 
It is essential to note that this paper does not explore the accuracy assessment of each data source.

\subsubsection{Trains}
The railway network across Europe exhibits a diverse and country-specific management infrastructure. Unfortunately, no open-source API or standardized pan-European platform would enable us to aggregate data seamlessly from various railway networks. Consequently, our approach relied on country-specific rail networks, focusing on Germany due to easier data accessibility.
We utilized web scraping techniques on the Bahn.Expert website to gather the necessary information. This platform uses the Deutsche Bahn APIs~\footnote{https://data.deutschebahn.com/dataset.groups.apis.html} to provide valuable insights into past and present data on trains and stations, particularly for Deutsche Bahn (DB) or German trains~\citep{bahnexpert}.

In our data collection process, our primary emphasis was on three prominent long-distance train categories in Germany --- Intercity (IC), EuroCity (EC), and Intercity-Express (ICE). This focus was chosen because these categories encompass most of the major cities in Germany and extend to significant cities in neighboring countries, including Amsterdam, Vienna, Basel, Brussels, and others, as illustrated in~\autoref{fig:european-cities}. 
We chose to focus on the data from DB, as it offers valuable insights into connections with major cities in Germany and neighboring countries. 
However, it’s important to note that our methodologies are adaptable and can be extended to incorporate data from other train providers across different European regions.

\begin{table*}[ht]
      
    \centering
    \caption{Summary of data sources and basic statistics for calculating emissions upon arrival at the destination along with the~\COtwoE~ values used for each category}
    \resizebox{\linewidth}{!}{\begin{tabular}[t]{llllllllllc}
       \toprule
       \multicolumn{1}{l}{\multirow{2}{*}{\textbf{Mode}}} & \multicolumn{1}{l}{\multirow{2}{*}{\textbf{Data Sources}}} & \multicolumn{1}{l}{\multirow{2}{*}{\textbf{Category}}} & \multicolumn{1}{c}{\multirow{2}{*}{\textbf{\begin{tabular}[c]{@{}c@{}}\# Unique\\ Routes\end{tabular}}}} & \multicolumn{3}{c}{\textbf{Distance (km)}} & \multicolumn{3}{c}{\textbf{Travel Time (hrs)}} &  \multicolumn{1}{c}{\multirow{2}{*}{\textbf{\begin{tabular}[c]{@{}c@{}}\COtwoE\\ (g/km)\end{tabular}}}} \\
       & & & & \textbf{Min} & \textbf{Mean} & \textbf{Max} & \textbf{Min} & \textbf{Mean} & \textbf{Max} & \\
       \midrule
       \multirow{4}{*}{Flights} & \multirow{4}{*}{Google Flights} 
       & Very Short Haul & 1,617 & 30.46 & 331.68 & 499.70 & 1.08 & 6.79 & 50.25 & 155 \\
       & & Short-Haul & 7,086 & 501.34 & 995.33 & 1,499.78 & 1.08 & 7.21 & 83.17 & 110 \\
       & & Medium-Haul & 7,450 & 1,500.42 & 2292.91 & 3,986.05 & 2.03 & 10.82 & 53.75 & 75 \\
       & & Long-Haul & 108 & 4,004.87 & 4,317.96 & 4,986.29 & 8.33 & 18.41 & 55.25 & 95 \\
       \midrule
       \multirow{2}{*}{Driving} & Google Maps API & -- & 4,718 & 47.24 & 851.51 & 2,440.34 & 0.77 & 10.14 & 35.24 & \multirow{2}{*}{96}\\
       & OSRM API & -- & 5,028 & 43.67 & 917.03 & 2,967.52 & 0.72 & 11.08 & 37.67 \\
       \midrule
       \multirow{3}{*}{Trains} & \multirow{3}{*}{Deutsche Bahn} & ICEs & 104 & 116.17 & 524.75 & 849.67 & 2.10 & 6.85 & 10.88 & \multirow{3}{*}{24} \\
       & & ICs & 144 & 24.51 & 236.86 & 672.77 & 0.68 & 3.66 & 9.27  \\
       & & ECs & 101 & 30.77 & 350.17 & 779.25 & 0.68 & 5.95 & 14.15  \\
       \bottomrule
    \end{tabular}}  

     \label{tab:transportation-summary}
\end{table*}

\subsection{Estimation of Emissions from Transportation Modes} \label{section: ghg estimations}

\rebuttal{Transportation emissions are a major contributor to environmental harm, driving climate change through the release of greenhouse gases (GHGs) such as carbon dioxide (\tunar{\CoTwo}), methane (CH\textsubscript{4}), and nitrous oxide (N\textsubscript{2}O)~\citep{lashof1990relative}. 
Among these, \tunar{\CoTwo} is the most prevalent greenhouse gas emitted by human activities, both in quantity and its overall impact on global warming. 
While \tunar{\CoTwo} is often used as a general term for greenhouse gases, this paper employs \textit{"carbon dioxide equivalent"} (\tunar{\COtwoE}) to encompass the broader spectrum of emissions and their combined impact~\citep{brander2012greenhouse}.}

While there were discrepancies among individual studies regarding the exact emissions per kilometer, the overall consensus indicates that short-distance flights have a higher environmental impact. In contrast, trains and public transport significantly reduce \tunar{\COtwoE}~emissions. Our modeling employs an average approach, which is in line with the above consensus.~\autoref{tab:transportation-summary} presents a summary of values used for~\COtwoE~calculations across various transportation modes.
The subsequent sections provide a detailed exploration of the emission calculation process and our assumptions for each mode of transportation.

\subsubsection{Flights} 

Google Flights calculate emissions per person using various factors, including the GCD between origin and destination airports, aircraft type, fuel burn, and flight occupancy following the Tier 3 methodology for emission estimates outlined by the European Environment Agency (EEA)~\citep{google_travel_impact_model}. However, the calculation results in a per-passenger~\COtwoE~ contribution, which can be misleading compared to the overall fuel consumption for the entire journey. To facilitate a fair comparison with alternative transportation modes, such as rail and driving, we adopt a distance-based estimation model proposed by~\citet{graver2019emissions} for standard economy class flights. 
\autoref{tab:transportation-summary} displays the distinct~\COtwoE~values applied to three flight categories, categorized according to the covered distance.
We also add an extra 9\% correctional adjustment factor to the great-circle distance to account for delays and indirect flight paths, as noted by~\citet{defra_actonco2_calculator}.

\subsubsection{Driving}

Estimating driving emissions involves several factors: elevation, car model, fuel type, car size, number of occupants, and traffic conditions~\citep{ghosh2020eco}. 
To calculate the~\COtwoE~for the given route, we utilize the per-kilometer emission estimation from \citet{statista_carbon_footprint}, set at 96 grams per kilometer.
The Google data also included fuel consumption estimates, allowing us to derive the~\COtwoE~values. 
We also compute the fuel consumption-based~\COtwoE~for the Google data, with an emission rate of 2.3 kilograms of~\tunar{\COtwoE}~per liter of gasoline~\citep{hilali2019contribution}.
The mean absolute difference percentage of 10.90 between \COtwoE~values calculated from per-kilometer distance and those derived from fuel consumption estimates for the Google data indicates moderate variability or discrepancy between the two methods. 
To maintain simplicity and ensure standardization in the estimation calculation, we adopt distance-based estimates as listed in~\autoref{tab:transportation-summary} computed for the minimum distance returned by either the Google or the OSRM data.

\subsubsection{Trains}

Much like other modes of transportation discussed earlier, train emissions can vary depending on the type of fuel used. In Europe, where electric trains are prevalent, emissions are considerably lower than diesel-powered counterparts. However, reported values exhibit discrepancies even within Europe's predominantly electric rail network. For instance, Statista UK cites 41 grams of~\COtwoE~ per kilometer~\citep{statista_carbon_footprint}, while Deutsche Bahn reports 32 grams of~\COtwoE~per passenger kilometer~\citep{statista_bahn}. 
Our estimations are based on values obtained for trains from~\citet{larsson2023methodology}, specifying 24 grams of~\tunar{\COtwoE}~per kilometer. We acknowledge the inherent challenges in establishing a universally accurate emission figure for this context.

\subsection{Estimating the Transportation Trade-offs} \label{section: transportation tradeoff}

The values representing the trade-off between emissions, travel time, and cost are relevant in pinpointing users with stronger pro-environmental attitude and formulating effective policies. 
\rebuttal{Following the recommendation by~\citet{aziz2014exploring},
we define $\EmissionScorei$ for city $\Cityi$ as the \textit{emissions trade-off index} to compute the trade-off among travel time ($TT$), \tunar{\COtwoE}~emissions ($EM$), and cost across all available transportation modes when selecting a trip. 
Our formulation is detailed as follows:}
\begin{equation}
    \label{eq: emission score}
    \EmissionScorei = \alpha_{TT} \cdot \tau_{TT}({\Cityi}) + \alpha_{EM} \cdot \tau_{EM}({\Cityi}) + \alpha_{Cost} \cdot \tau_{Cost}({\Cityi})\  
   \end{equation}
In this equation, $\alpha_{j}$ represents the weight associated with each element, $\tau_{j}({\Cityi})$ signifies the normalized trade-off index associated with each element for the city $\Cityi$, where $j \in \{TT, EM, Cost\} $.

\rebuttal{
Estimating transportation costs is challenging due to their variability, which is influenced by external factors such as booking timing and method. Therefore, we used estimation methods for this study.}
In the case of calculating flight costs to reach a destination, we leverage per kilometer estimates provided by~\citet{rome2rio2018} for the top 200 airlines and their respective domestic and international flights. Our approach involves mapping the airline's per kilometer price, specifically when booked four weeks in advance, to the international and domestic categories based on whether the destination is within the same or a different country, respectively.  We acknowledge the challenges in mapping costs for multi-carrier airlines, limiting our current model to direct flights or those with a single airline. 
We adopt the estimation provided by~\citet{euronews2023} for train travel, setting the cost at 0.14 euros per kilometer for tickets booked four weeks in advance. When it comes to driving, we determine costs based on estimations of the average cost per kilometer of fuel for the country where the journey originates. To achieve this, we use data from~\citet{alternativeFuelsObservatory} for country-specific fuel price estimations.
While these estimates may not be entirely precise, they serve as a helpful tool for cost modeling.

This paper focuses on data involving up to three transportation modes between cities. We compute travel time, emissions, and cost trade-offs for each trip across all available transportation modes. To ensure consistency, we normalize time (in hours), emissions (in kilograms), and costs (in euros) across all modes between values 0 and 1 using min-max normalization~\citep{patro2015normalization}. This normalization also yields relative values, allowing comparisons across different transportation modes.
Mathematically, the normalized trade-off for each element can be calculated as follows:
\begin{equation}
    \label{eq: min_max_norm emissions}
    N_j({\Cityi}) = \frac{V_{j}^{(i)} - \min(V^{(i)})}{\max(V^{(i)}) - \min(V^{(i)})}
\end{equation}
Where 
$V^{(i)} = \{V_{TT}^{(i)}, V_{EM}^{(i)}, V_{Cost}^{(i)}\}$ is the set of factors across all modes of transportation involved in the \tunar{\textit{emissions trade-off index}} for a trip.
In this context, each element of $\tau_j({\Cityi})$ varies between 0 and 1. Zero signifies the most favorable alternative, whereas one indicates the least favorable one.
We learn the weights $\alpha_{TT}$, $\alpha_{EM}$ and $\alpha_{Cost}$ from the user study explained in~\autoref{section: user_study}.

By examining trade-off values, we can gain insights into user behavior and formulate effective policies, such as implementing strategies to enhance travel time on routes with lower emissions and higher costs. 
The lower the~$\EmissionScorei$~for a city~$\Cityi$~, the less damaging it is to visit that city from the users' point of origin and thus more \textit{fair} from a societal perspective.
However, it is important to recognize that the interpretation is context-sensitive, as travel time, emissions, and costs are distinct variables that cannot be interchanged.
The model informs us explicitly about the trade-off involved in travel decision-making for sampled users, considering factors like trip duration, emission, and context.

\subsection{Summary} \label{section: ghg summary}

\rebuttal{This section outlines the methodology used to collect data and estimate \tunar{\COtwoE}~emissions for transportation to various European cities concerning a user's starting location. 
We collected data for three transportation modes: flights from Google Flights, driving from Google Maps Routes API and OSRM API, and train connections for major German cities and their neighboring countries from Bahn.Expert website. 
While this study focused on European data sources, the methods employed are adaptable and can be extended to incorporate data from other providers worldwide.}

\rebuttal{To account for the~\tunar{\COtwoE}~emissions for each mode of transport, we used a simplified distance-based model utilizing the concept of  \textit{"emissions trade-off index"} to evaluate trade-offs among travel time,~\tunar{\COtwoE}~emissions, and cost across different modes of transport. 
A lower index indicates less environmental impact from visiting a city from the user's origin, enhancing societal fairness. 
Analyzing these trade-off values offers insights into user behavior, informing potential policy strategies to optimize routes with lower emissions and higher costs. 
While our model sheds light on travel decision-making trade-offs, future iterations could incorporate additional factors contributing to~\tunar{\COtwoE}~emissions during city trips, such as accommodation and food consumption.}

\section{Tourist Destination Popularity} 
\label{section: popularity}

\rebuttal{Cities often struggle with the negative impacts of tourism, such as elevated housing prices, intensified traffic, and increased congestion, leading to resident dissatisfaction and anti-tourism sentiment~\citep{camarda2003environmental, Gowreesunkar2020, seraphin2018over}. 
To combat this, "de-tourism" initiatives aim to redirect tourists towards less crowded, alternative destinations. Promoting these hidden gems through official channels and collaborative social marketing helps distribute tourism more evenly and alleviates pressure on popular spots~\citep{santropez2020abc}. However, destination popularity, largely driven by online presence, plays a significant role in tourist decision-making. Search engine activity, social media engagement, and user-generated content like reviews and captivating visuals contribute to a location's attractiveness~\citep{weng2022local}. 
Research indicates that a substantial 83\% of tourists utilize Google Images for destination-related searches before embarking on their travels and these images are pivotal in shaping tourists' perceptions of a destination and influencing their travel decisions~\citep{pan2007travel, tasci2007destination}.
Therefore, when recommending sustainable travel destinations from a particular city, it is important to go beyond the popular options and also discover the lesser-known cities or hidden gems that are not frequented by many tourists. This approach helps to balance the tourist load more evenly among cities.
}

In our study, we assess popularity across three dimensions --- the prevalence of city images searched on Google (GT), the number of points of interest (POI) at a destination, and various forms of user-generated content (UGC), such as reviews and photos.
These components serve as proxies for popularity, and a strong correlation between the number of reviews and attractions indicates a highly popular destination. The collected data is normalized to derive a \textit{popularity index} for each city.
The city's \textit{popularity index} can be incorporated into a recommender system, aiding users in decision-making and promoting the selection of destinations with lower popularity.

\subsection{Data Gathering} \label{subsection: datasets popularity}
To estimate the popularity of a city, we gathered data from two prominent sources --- Tripadvisor and Google Trends. 
The sections below explore the details of the data-gathering process.

\begin{table*}[htbp]
    \caption{Table summarizing different data sources and their basic statistics used for calculating the \textit{popularity index} for a city}
    \label{table:data_sources_popularity}
    \resizebox{\linewidth}{!}{\begin{tabular}[t]{lllllll}
        \toprule
        \theadb{1}{Data Sources} & \theadb{2}{Attributes} & \theadb{4}{Statistics} \\
        \cmidrule{4-7}
        &  &  &  Min & Mean & SD & Max \\
        \toprule
        \multirow{4}{*}{Tripadvisor} & POI & \# Attractions &  5.0  &    683.80       &   1,338.59     &  8,999   \\
        & \multirow{3}{*}{UGC} & Total \# Reviews \& Opinions  &   217  &  302,935.5  &  853,071.2  &   7,099,844  \\
        &  & \# Attraction Reviews   &  0  & 74,058.54  & 212,555.9  & 1,795,447  \\
        &  & \# Attraction Photos   &   0 & 56,214.04 &  140,800.1  &  1,019,360    \\
        \midrule
        Google Trends & GT & Images  &  0 & 13.70 & 5.90 & 100   \\
        \bottomrule
    \end{tabular}}
\end{table*}

\subsubsection{Tripadvisor}
Tripadvisor is a popular online platform aggregating user-generated reviews and ratings for travel-related entities, such as accommodations, restaurants, and attractions.
To estimate the popularity of cities, we utilized web scraping techniques on the Tripadvisor platform to gather key metrics such as the total number of reviews, number of attractions (POI), reviews on attractions, and photos of attractions for each of the 200 European cities that we had gathered in~\autoref{section: extracting cities}.
\rebuttal{Major tourist cities such as London, Paris, and Rome exhibit the maximum number of attractions, reviews, opinions, and photos, a finding consistent with expectations.}
~\autoref{table:data_sources_popularity} summarizes the basic statistics of the Tripadvisor data.

The data reveals a striking similarity between attraction review lists and photo counts. 
To validate our hypotheses, we conducted a correlation analysis examining the relationship between the overall reviews and opinions, the number of attraction reviews, and the number of attraction photos at each destination. The analysis revealed an exceptionally strong correlation, exceeding 0.90. Additionally, a T-Test~\citep{eisenhart1979transition} was performed to assess the significance of this correlation, and the results confirmed its statistical significance.
Therefore, we consider a combined count of reviews and opinions for a particular city for our \textit{popularity index}, serving as a proxy for all the user-generated content elements.

\subsubsection{Google Trends}

We used Pytrends~\footnote{https://pypi.org/project/pytrends/} to collect weekly data from Google Trends (GT) within the travel category for the year 2022. Focused worldwide and on English-language search results, GT normalizes search data to facilitate comparisons between terms. This normalization involves dividing each data point by the total searches within its corresponding geography and time range, ensuring relative popularity comparisons and preventing biases toward regions with higher search volumes. The resulting values are scaled from 0 to 100, reflecting a topic's proportion to all searches. It's important to note that regions with the same search interest for a term may not always share the same total search volumes~\citep{google_trends}.

We conducted a correlation analysis between the GT and Tripadvisor data to gain deeper insights into the data, focusing specifically on POI and UGC components for the respective cities. Surprisingly, the analysis revealed a notably low correlation but was statistically significant. This suggests that the patterns in Google search trends for cities, as measured by image searches, do not strongly align with the popularity of attractions and user-generated content on Tripadvisor. 
Therefore, we treat GT as a distinct entity in our \textit{popularity index} estimation, recognizing its divergence from other indicators.%
\begin{figure}[h]
    \centering
    {\includegraphics[width=0.9\textwidth]{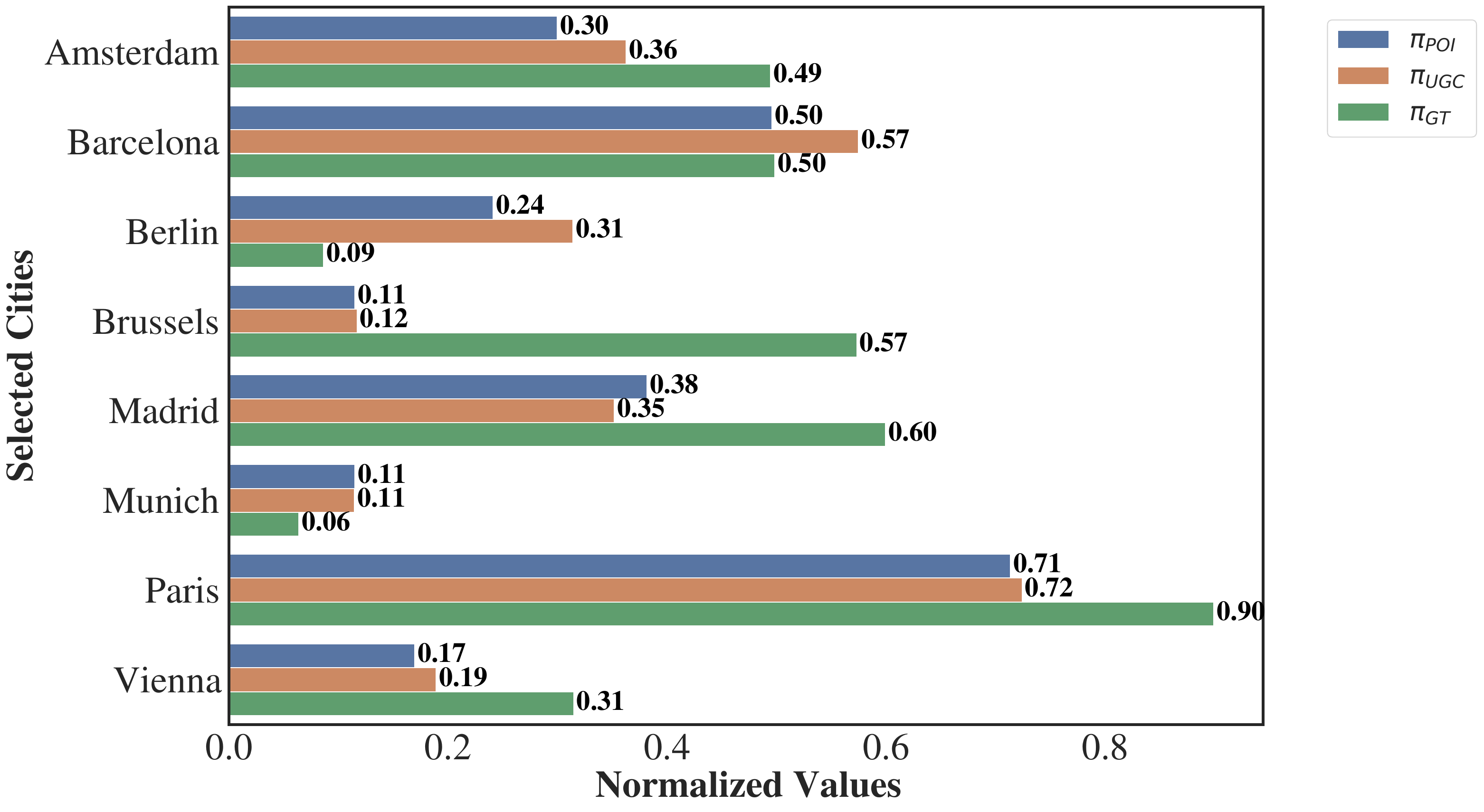}}
    \caption{Bar plot showing the normalized values of the \textit{popularity index} components for selected cities}
    \label{fig: popularity selected cities}
\end{figure}

\subsection{Estimating Destination Popularity} 

Quantifying a city's popularity is complex due to its multifaceted nature. This paper proposes a method to define city popularity based on metrics derived from Tripadvisor and the GT data as explained above in~\autoref{subsection: datasets popularity}. 
The \textit{popularity index} $\popScoreCi$ of a city $\Cityi$ is expressed as a weighted sum of various popularity components —- $\pi_{POI}$, $\pi_{UGC}$, and $\pi_{GT}$ denoting the points of interest, number of reviews and opinions available on Tripadvisor and google trends index for the last one year respectively for a city~$\Cityi$.
To ensure consistency, we employ min-max normalization~\citep{patro2015normalization}, as depicted in~\autoref{eq: min_max_norm emissions}, to standardize all component values within the range of 0 to 1. 
~\autoref{fig: popularity selected cities} displays the normalized values of various components constituting the \textit{popularity index} for a chosen group of cities. The contrast in popularity is evident, with larger cities like Paris and Barcelona grappling with overtourism, while Munich and Berlin, comparatively less popular, showcase a distinct difference.
Based on this, we define the \textit{popularity index} $\popScoreCi$ for the city~$\Cityi$~as follows:

\begin{equation}
    \label{eq: popularity score}
    \rho(c_{i}) = \beta_{POI} \cdot \pi_{POI} + \beta_{UGC} \cdot \pi_{UGC} + \beta_{GT} \cdot \pi_{GT}
\end{equation}
Where $\beta_j$ for $j \in \{POI, UGC, GT\} $ are the weights assigned to each component of the \textit{popularity index}.
We derive the weights, $\beta_j$, for the \textit{popularity index} components through a user study, as detailed in~\autoref{section: user_study}. This process involves determining the quantitative contributions of factors within the popularity elements to the overall \textit{popularity index}, relying on user preferences and perceptions. By adopting this user-driven approach, the assigned weights for each component reflect user opinions and behaviors, thereby incorporating a human-centered dimension into the calculation of city popularity.
We aim to recommend destinations with a lower \textit{popularity index} to foster a balanced distribution of tourist traffic, even in less popular yet attractive cities, thus mitigating overtourism at popular destinations.

\subsection{Summary} \label{section: popularity summary}

\rebuttal{In this section, we investigate destination popularity and its influence on tourist behavior, particularly in addressing the challenge of overtourism. To recommend sustainable destinations, TRS must balance well-known options with lesser-known gems, distributing tourist traffic more evenly among cities.}
\rebuttal{We utilized Tripadvisor and Google Trends data to quantify city popularity based on user reviews, points of interest, and search trends. Our proposed method assigns a \textit{popularity index} to the cities using metrics from these sources and recommends destinations with a lower \textit{popularity index}. 
This allows us to prioritize recommendations for less popular cities while considering their attractiveness, aiming to promote a more balanced and sustainable tourism ecosystem.}

\section{Seasonal Destination Demand} \label{section: seasonality}

\rebuttal{The personalized recommendation algorithms on online platforms often prioritize specific destinations, leading to a high concentration of tourists during certain seasons while overlooking less-visited places~\citep{Gowreesunkar2020}.}
TRS can intervene in this issue by redirecting tourists to less crowded destinations and avoiding peak seasons, thereby ensuring a consistent distribution of tourists throughout the year across all seasons.
\rebuttal{Cities can exhibit a variety of seasonal touristic patterns, including single peaks, dual peaks, and consistent year-round visitation, with causes attributed to both natural factors like climate and institutional factors such as holidays and cultural practices~\citep{butler1998seasonality, corluka2019tourism,seasonality2020croatia}. Our objective is to assign a seasonal demand index to each city for a given month, providing an estimation of its appeal to tourists. The aim is to recommend destinations with a low \textit{seasonality index}, ensuring a consistent tourist presence throughout the year or balancing tourist loads across different destinations.
}

\rebuttal{When generating recommendations, the \textit{seasonality index} provides complementary information to the \textit{popularity index}.}
While the \textit{popularity index} offers an aggregated view of a destination's appeal over the entire year, the \textit{seasonality index}, on the other hand, provides a more nuanced, month-by-month analysis. 
This distinction is essential for travelers planning their trips. 
Although helpful, the \textit{popularity index} might not fully capture a destination's unique characteristics or visitor trends in a specific month. In contrast, with its monthly granularity, the \textit{seasonality index} offers a more accurate representation of what a traveler can expect during their chosen travel period.

\subsection{Data Gathering}
To gauge the seasonal variations in tourist activity, we explore monthly visitor counts and bednight statistics from TourMIS and financial seasonality indicators, such as the average daily rates (ADR) from Airbnb. The ensuing sections elaborate on these data-gathering processes.

\begin{table*}[h]
    \centering
    \caption{Table summarizing different data sources and their basic statistics for calculating the destination's \textit{seasonality index}}
    \label{table:data_sources_seasonality}
    \resizebox{\textwidth}{!}{
    \begin{tabular}{lllcllllll}
    \toprule
    \multicolumn{1}{c}{\multirow{2}{*}{\textbf{Data Source}}} & \multicolumn{2}{c}{\multirow{2}{*}{\textbf{Attributes}}} & \multicolumn{5}{c}{\textbf{Statistics}} \\ \cmidrule{4-8} 
    \multicolumn{1}{c}{} & \multicolumn{2}{c}{} & \multicolumn{1}{l}{\# Cities} & Min & Mean & SD & Max \\ \toprule
    \multirow{2}{*}{TourMIS} & \multirow{2}{*}{NFIs} & AVC & 64 & 489 & 192,125.56 & 293,783.60 & 2,188,497 \\ 
 &  & BN & 69 & 1,163 & 415,708.25 & 675,550.57 & 5,260,073 \\      
    Airbnb & FI & ADR & 45 & 68.09 & 316.88 & 418.63 & 2013.13 \\       
    \bottomrule
    \end{tabular}
    }
\end{table*}

\subsubsection{TourMIS}

TourMIS~\footnote{https://www.tourmis.info}, a tourism marketing information system, offers complimentary and electronically accessible market research data to aid management decisions. Supported by the regional, national, and international tourist industry, TourMIS provides up-to-date tourism statistics and analyses, including arrivals and bed nights, for informed decision-making~\citep{tourmis}. The monthly arrival visitor count (AVC), encompassing both foreign and domestic data, and the number of bed nights (BN) for European cities in 2022 are considered in our analysis.
The dataset includes information for 65 cities regarding AVC and 70 cities for BN, with an overlap of 63 cities. Leveraging this data, we estimate the footfall in cities for respective months. 

Exploratory data analyses on the TourMIS data revealed that a minimum AVC of 489.0 was recorded in January for the city of Eisenstadt, while Paris recorded a maximum AVC of 2,188,497 in July as evident from~\autoref{table:data_sources_seasonality}. Similarly, Eisenstadt, Austria, documented the minimal number of bednights for January, while Paris still accounted for the maximum number in July. These insights illuminate the dynamic nature of tourism, showcasing the fluctuating visitor counts and bednights across various months and cities. Notably, the data suggests heightened touristic activity during the summer compared to winter.

We performed correlation analyses on the bednights (BN) and monthly AVC data from TourMIS. 
The results unveiled a consistently strong correlation (>0.98) for each month among cities where data for both variables were available.
To determine if this correlation trend holds for the entire population of cities, we carried out a T-Test~\citep{eisenhart1979transition} with a significance level (p) of 0.05. 
The test yielded significant results, indicating sufficient evidence to conclude that the correlation significantly differs from zero in the population. 
Therefore, for our subsequent calculations, we exclusively consider the AVC numbers.
It's important to acknowledge that our approach utilizes absolute figures for AVC without normalizing them based on the size of the cities. 
Although attempts were made to normalize city sizes, the outcomes were heavily skewed towards small cities.
Our method accurately reflects the seasonal demands of smaller cities, which often feature numerous attractions. 

\subsubsection{Airbnb}

\begin{figure}[h]
    \centering
   
    \begin{subfigure}{.3\textwidth}
        \includegraphics[width=\linewidth]{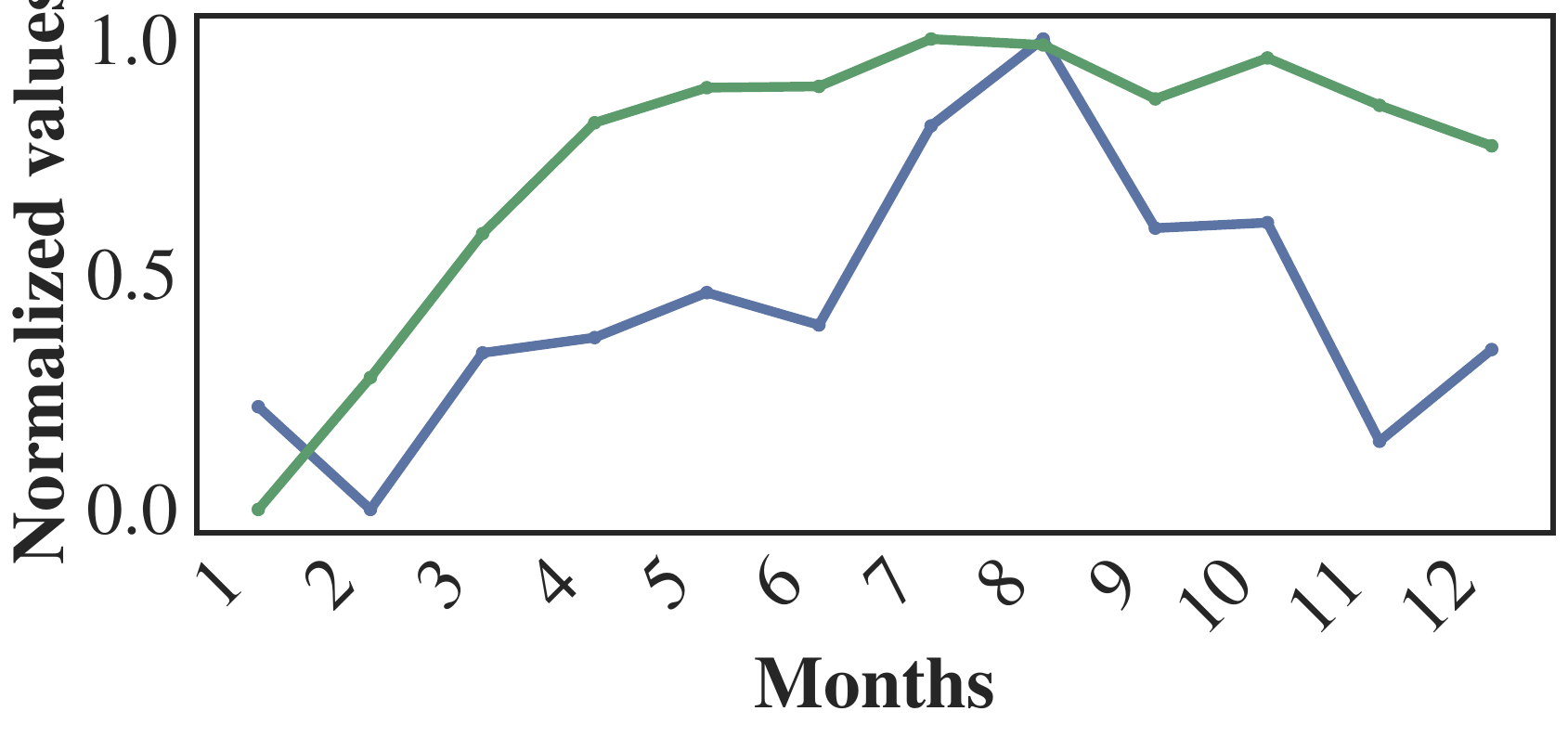}
        \caption{Amsterdam}
    \end{subfigure}
    \begin{subfigure}{.3\textwidth}
        \includegraphics[width=\linewidth]{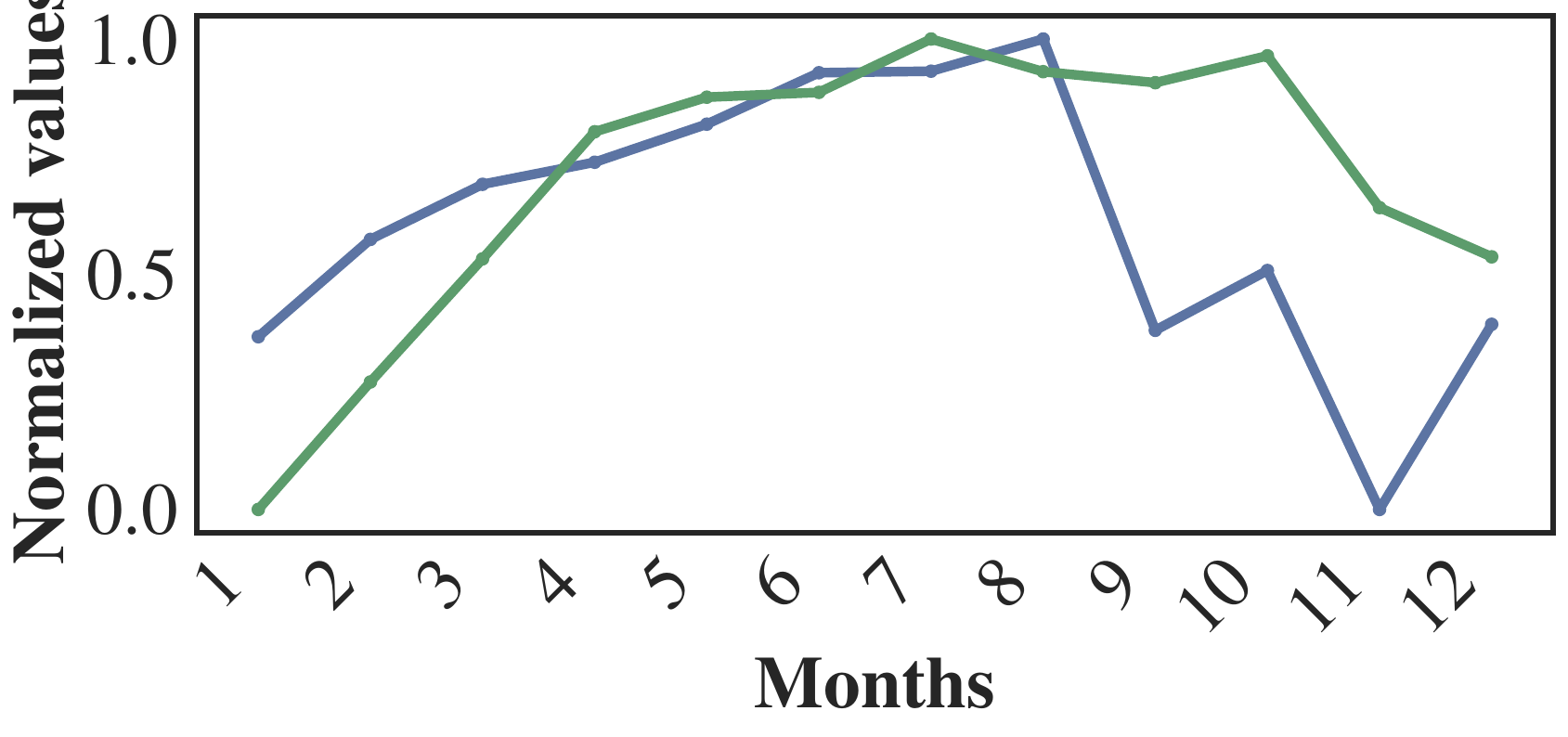}
        \caption{Barcelona}
    \end{subfigure}
    \begin{subfigure}{.3\textwidth}
        \includegraphics[width=\linewidth]{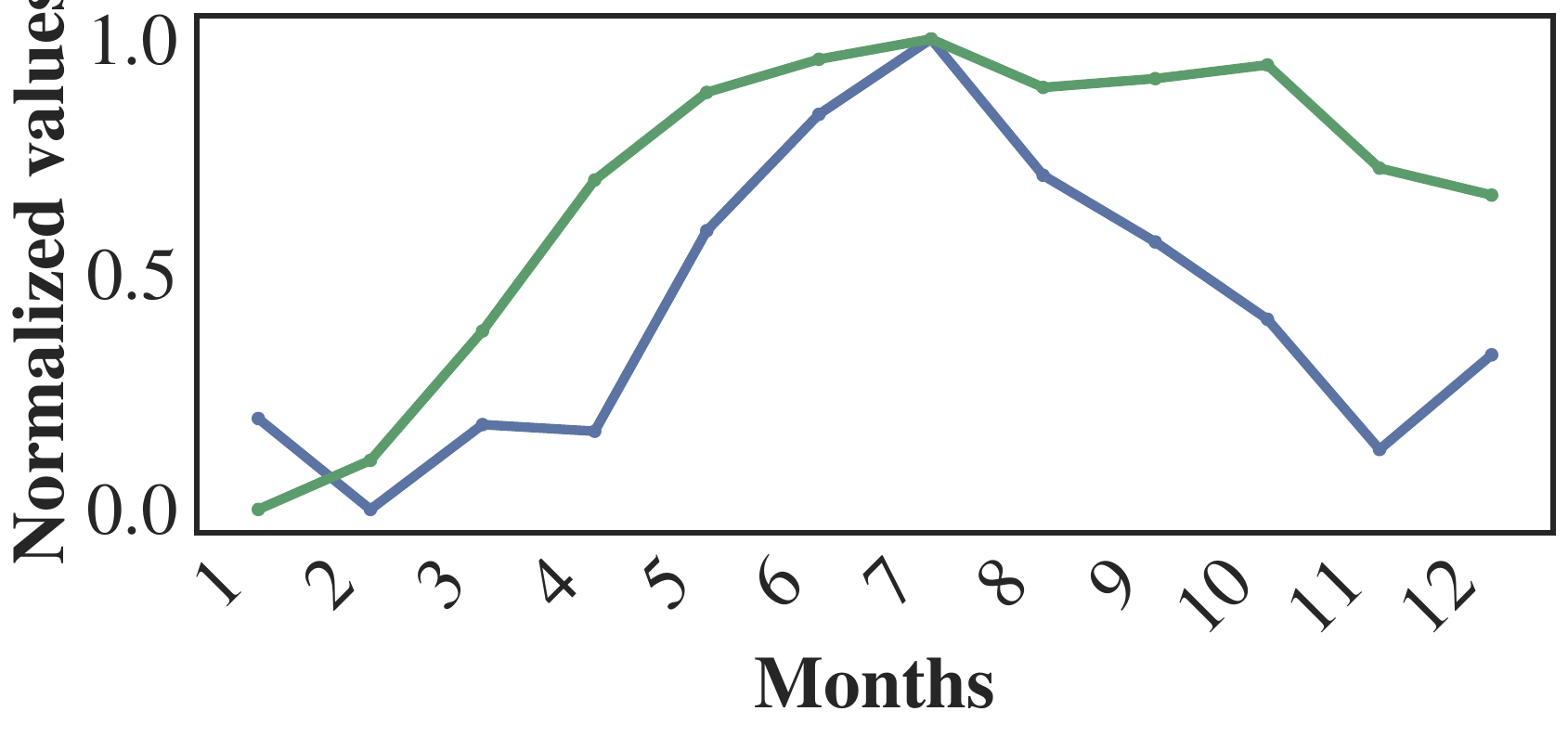}
        \caption{Berlin}
    \end{subfigure}
    \begin{subfigure}{.3\textwidth}
        \includegraphics[width=\linewidth]{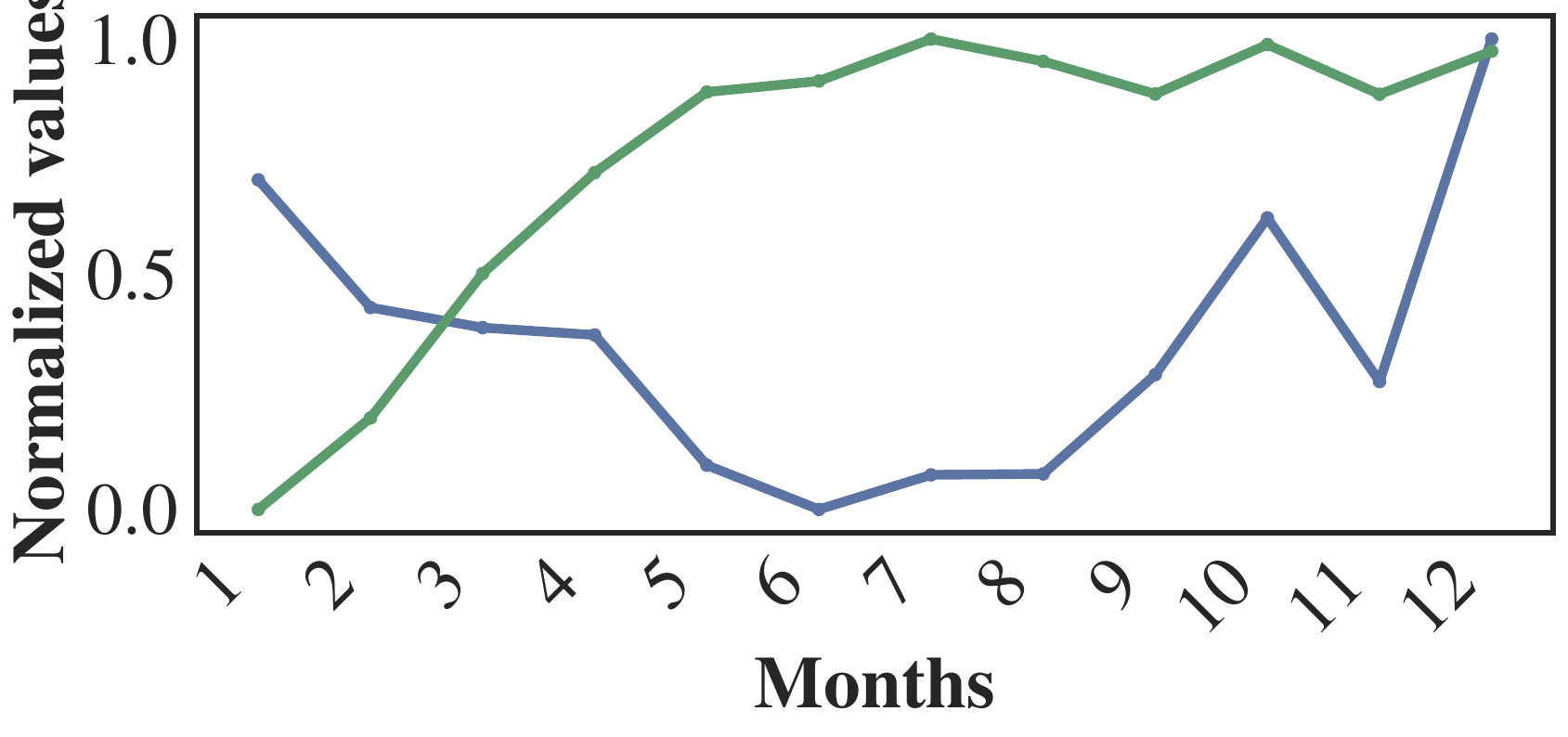}
        \caption{Brussels}
    \end{subfigure}
    \begin{subfigure}{.3\textwidth}
        \includegraphics[width=\linewidth]{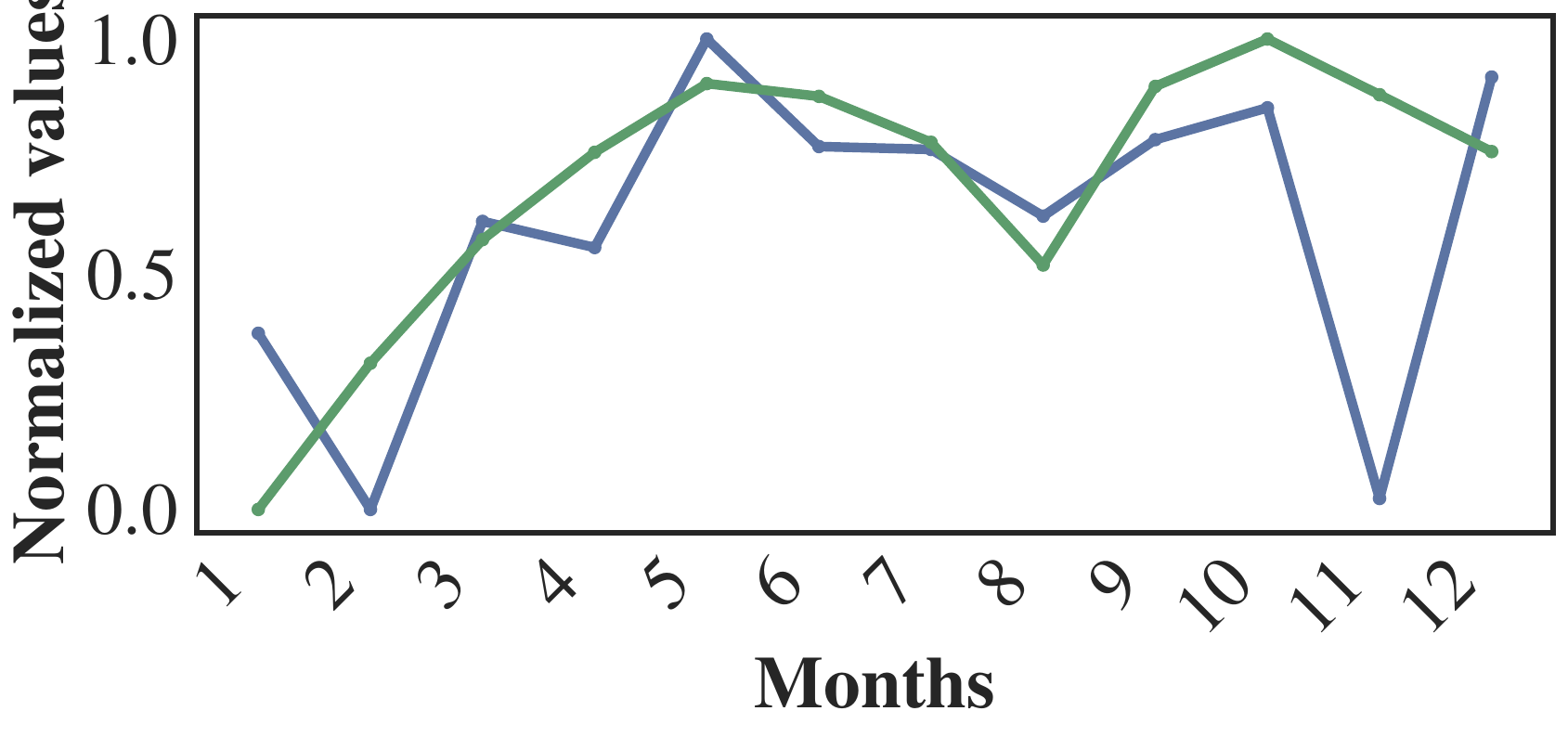}
        \caption{Madrid}
    \end{subfigure}
    \begin{subfigure}{.3\textwidth}
        \includegraphics[width=\linewidth]{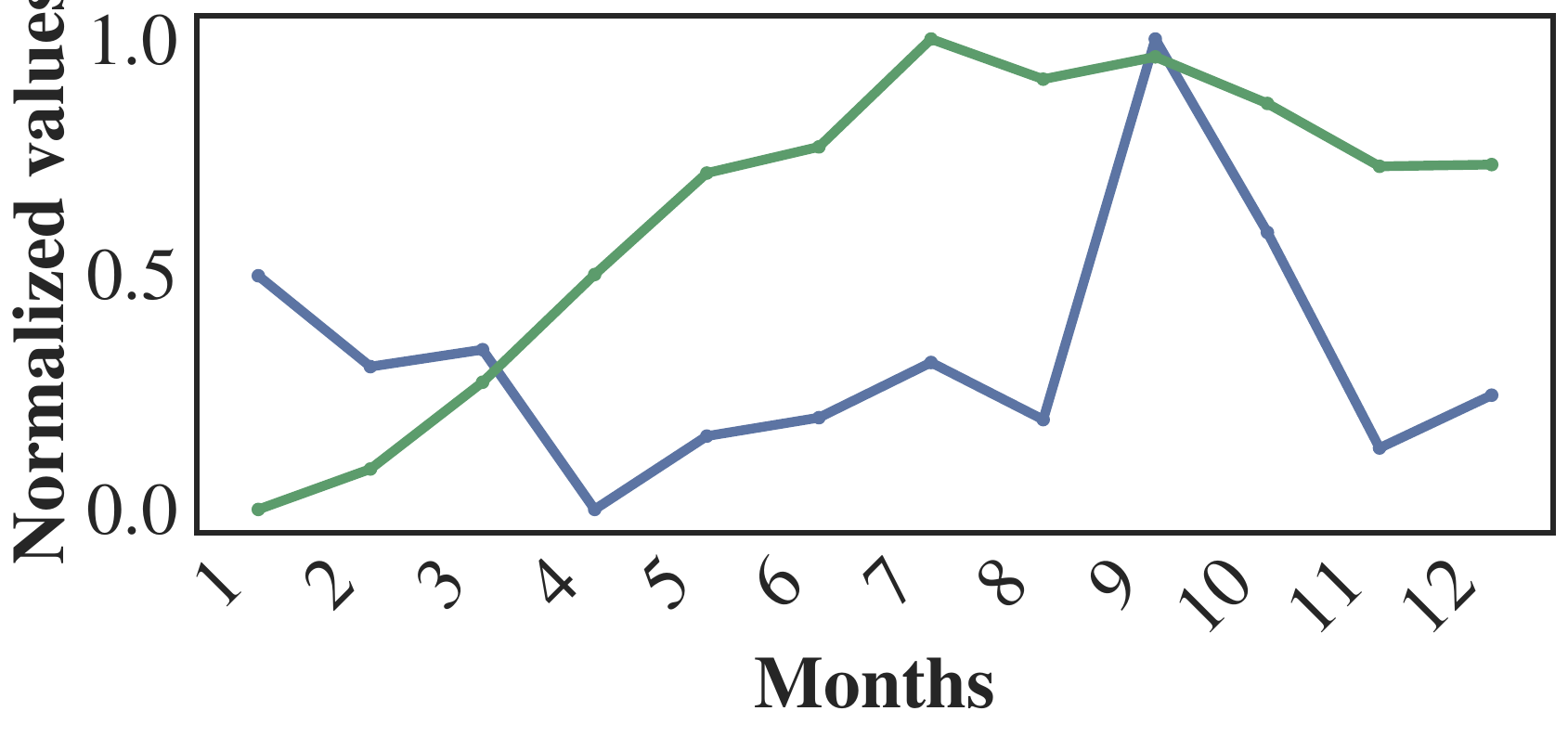}
        \caption{Munich}
    \end{subfigure}
    \begin{subfigure}{.3\textwidth}
        \includegraphics[width=\linewidth]{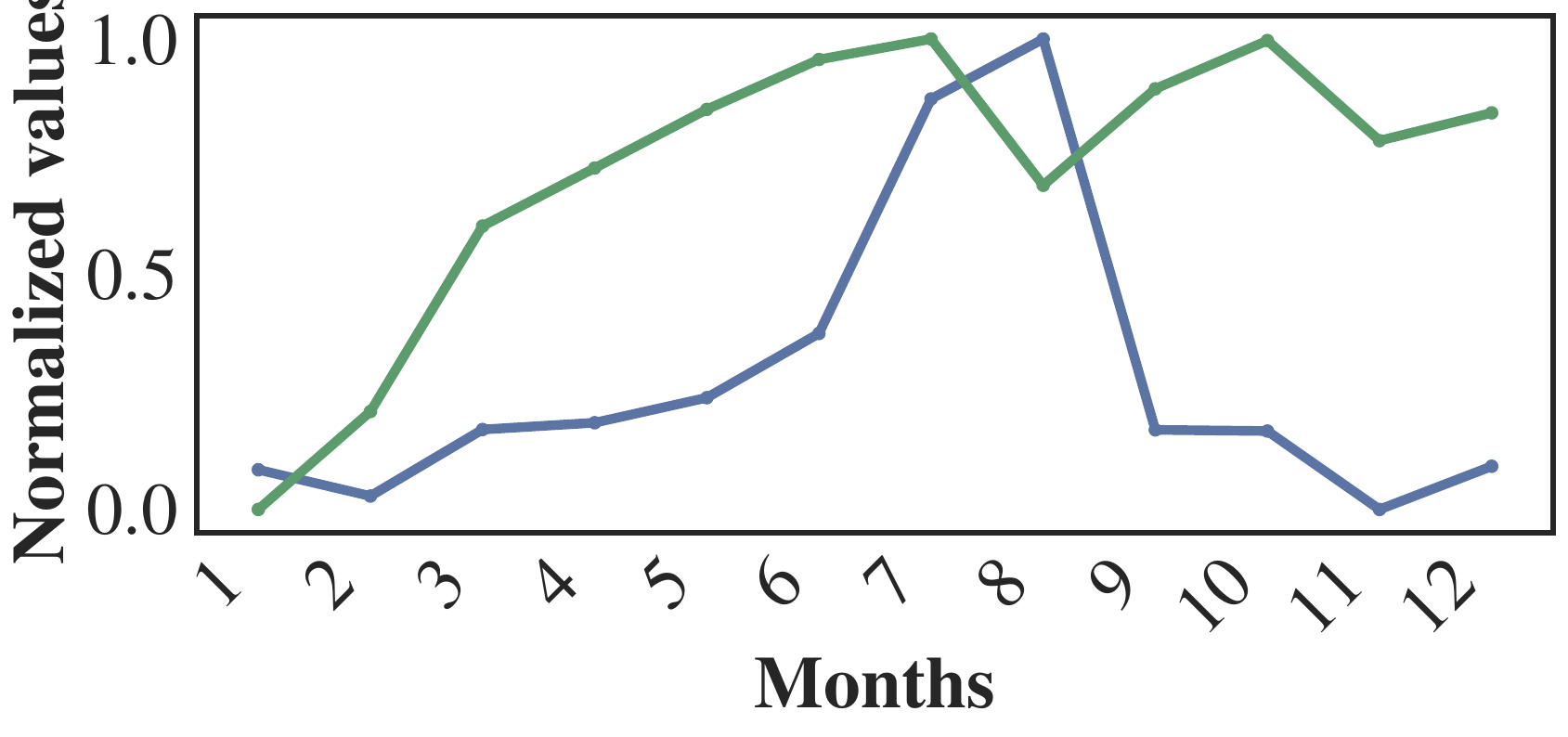}
        \caption{Paris}
    \end{subfigure}
    \begin{subfigure}{.3\textwidth}
        \includegraphics[width=\linewidth]{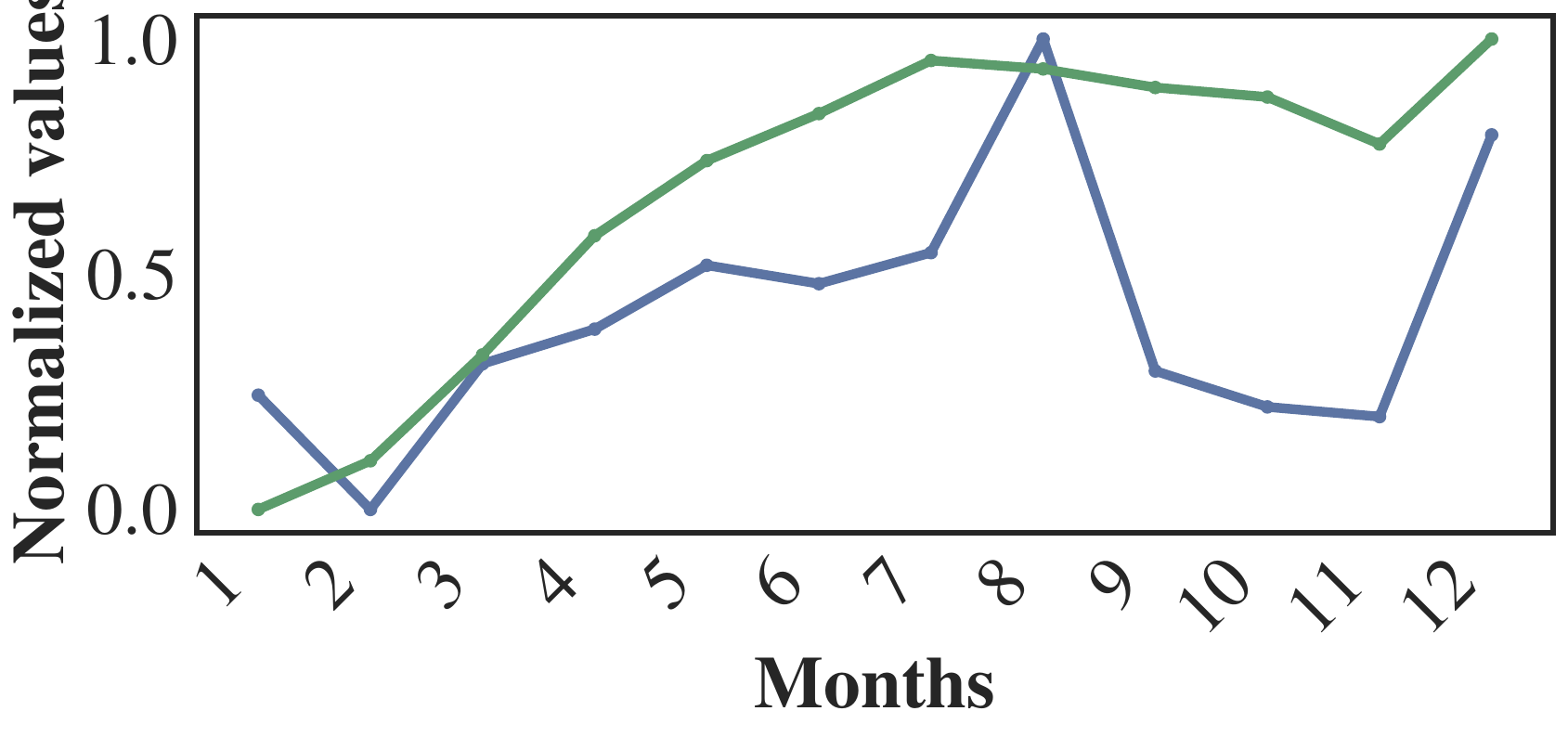}
        \caption{Vienna}
    \end{subfigure}
    \begin{subfigure}{.3\textwidth}
        \includegraphics[width=\linewidth]{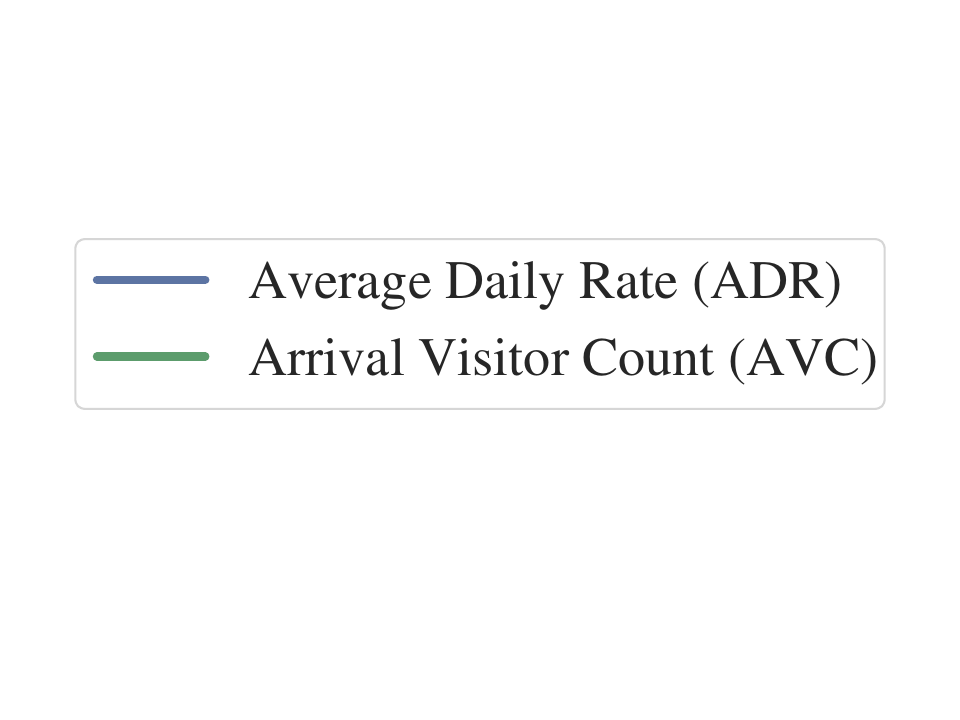}
    \end{subfigure}
    \caption{Visualization of the normalized monthly arrival visitor count (AVC) represented in green and monthly average listing price on Airbnb in blue for selected cities}
    \label{fig:adr_vs_avc_selected_cities}
\end{figure}

To quantify the impact of financial indicators like average daily rate (ADR) on seasonality, we leverage the \texttt{calendar.csv} dataset sourced from Inside Airbnb~\footnote{http://insideairbnb.com/get-the-data}. 
This dataset includes details on the availability and daily pricing of all listed accommodations within a city. 
Our analysis focuses on the latest data from September 2023 and covers a one-year duration for 45 European cities.

A foundational exploratory analysis of the data is presented in~\autoref{table:data_sources_seasonality}. 
In February, the lowest ADR was observed in Riga, while Oslo reported the highest ADR in July. 
These findings affirm the presence of seasonal demand variations across months, with increased demand in the summer, and align with the economic disparities between the cities~\citep{statista-house-price-to-income-ratio}. 
~\autoref{fig:adr_vs_avc_selected_cities} illustrates the normalized monthly AVC in green and the monthly average listing price on Airbnb in blue for selected cities. 
\rebuttal{The two variables exhibit a similar trend in most cases, except for Brussels. This deviation in Brussels can be attributed to more business travelers~\citep{santos2018tourism}.}
Specific peaks in Munich's accommodation prices during September are representative of Oktoberfest, while generally, prices are elevated in the summer months, followed by a gradual decline in the winter months.

Despite these observed \rebuttal{patterns}, the correlation coefficients for each month across all cities were negatively correlated and statistically insignificant. Consequently, we opted to include the ADR values of the cities for each month as a separate component in our analysis of the financial indicators of tourism demand.

\subsection{Estimating Seasonal Demand}

In literature, the Gini coefficient stands out as a widely employed tool for assessing tourism seasonality~\citep{seasonality2017iceland,seasonality2020croatia,ferrante2018measuring}. This coefficient offers distinct advantages, including its ability to consider distribution asymmetry, relative insensitivity to extreme values, and an indication of stability in the distribution of overnight stays within a single year~\citep{seasonality2020croatia}.
In our paper, the Gini coefficient serves as a numerical metric quantifying the level of inequality in the distribution~\citep{gini1921measurement}. Derived from the Lorenz curve, which illustrates the cumulative frequency of ranked observations starting from the lowest number, the Gini coefficient provides a comprehensive measure of the destination's demand at a particular time of the year.
The analytical formula frequently used for Gini coefficient calculation, applied in this paper, is expressed as~\citep{gastwirth1972estimation,seasonality2017iceland}:

\begin{equation}
    \label{eq: gini equation}
    G = \frac{2}{n} \sum_{i=1}^{n} (x_{i} - y_{i})
\end{equation}

where \begin{equation*}
    \begin{split}
        & n = \text{the number of fractiles, months, weeks, days, or other units} \\
        & x_{i} = \text{the rank of fractiles, for example, } \\
        & \quad \frac{1}{12}, \frac{2}{12}, \ldots \text{ when using months, or when using weeks } \\
        & \quad \frac{1}{52}, \frac{2}{52}, \ldots, \text{ or days } \frac{1}{365}, \frac{2}{365}, \ldots, \text{ etc. So } x_i = \frac{i}{n} \\
        & y_i = \text{the cumulated fractiles in the Lorenz curve} \\
    \end{split}
\end{equation*}

In the context of seasonal tourism demand, studies suggest that the average room price is one of the pivotal business performance indicators in the hotel industry~\citep{israeli2002star, o2011brands, pine2005performance}. We calculate the seasonality Gini index for a city using Gini coefficients derived from non-financial (NFI) and financial indicators (FI), as described by~\citet{seasonality2020croatia}. Non-financial indicators consist of monthly counts of arriving visitors (AVC) from TourMIS, while financial indicators involve the ADR computed from Airbnb listings.
The Gini coefficient values for each indicator span from \textit{zero} to \textit{one}, with \textit{zero} signifying a complete lack of seasonality and seems to be an active or equal distribution of volumes all year round. A Gini coefficient of \textit{one} indicates complete seasonality, i.e., the total volume is registered in one single month~\citep{seasonality2020croatia}.
Gini coefficients are computed annually for AVC, while monthly calculations are performed for ADR.

As illustrated in ~\autoref{fig:adr_vs_avc_selected_cities}, the monthly average prices of listings exhibit significant fluctuations depending on the city and month. Therefore, it is advisable to model these fluctuations daily to enhance the precision of our estimations. Similarly, if data is available daily or weekly granularity, the AVC numbers could be modeled at those levels for more accurate analyses.
The aggregated \textit{seasonality index}~$\sigmaCij$~ across all indicators for a city $\Cityi$  for month $j$ for can be calculated as follows:

\begin{equation}
    \label{eq: seasonality index}
    \sigmaCij = \Gamma_{AVC} \cdot G_{AVC}^{(i)} + \Gamma_{ADR} \cdot G_{ADR}^{(i)(j)}
\end{equation}

Where $\Gamma_{k}$ where $k \in \{AVC, ADR\}$ represents the weights as derived from the user study.
We aim to recommend cities with lower~$\sigmaCij$, ensuring a consistent tourist presence throughout the year or balancing tourist loads across different destinations.
\rebuttal{
After examining Pearson's correlation coefficient between the two data sources for a selected list of cities common in all datasets, we merged the non-financial indicator (AVC) and the financial indicator (ADR) to compute our \textit{seasonality index}. 
~\autoref{table: correlation_adr_avc} shows that while some cities like Amsterdam, Berlin, and Vienna exhibit a significant positive correlation, the correlation is not statistically significant for other cities, indicating an unclear relationship that cannot be relied upon as a single data source.
While ADR can capture short-term fluctuations in the tourism market, the AVC provides a broader perspective on tourism trends throughout the year. 
By integrating both, we achieve a balanced assessment of monthly seasonal demand for the cities.}

\begin{table*}[htb]
    \centering
    \caption{\rebuttal{Analysis of the relationship between Airbnb ADR and TourMIS AVC: Pearson's correlation coefficients and T-Test results for selected European cities}}
    \label{table: correlation_adr_avc}
    \resizebox{0.5\textwidth}{!}{
    \begin{tabular}{lcl}
        \toprule
        \textbf{City} & \textbf{\begin{tabular}[c]{@{}l@{}}Pearson's \\ Correlation \\ Coefficient\end{tabular}} & \textbf{\begin{tabular}[c]{@{}l@{}}T-Test \\ Results\end{tabular}} 
        \\ \midrule
        Amsterdam & 0.659 & Significant \\
        Barcelona & 0.458 & Not Significant \\
        Berlin & 0.746 & Significant \\
        Brussels & -0.312 & Not Significant \\
        Madrid & 0.515 & Not Significant \\
        Munich & 0.125 & Not Significant \\
        Paris & 0.317 & Not Significant \\
        Vienna & 0.611 & Significant \\
        \bottomrule
    \end{tabular}}
\end{table*}

We calculate the Gini coefficients for the AVC and ADR, presenting the results in~\autoref{table:gini-coefficients-combined}. 
Madrid exhibits the least seasonality in AVC, suggesting consistent demand throughout the year, while Munich registers the highest seasonality. 
Additionally, Brussels displays minimal seasonality in ADR (close to 0), and Munich exhibits maximum ADR seasonality in September and October. The heightened seasonality in Munich's AVC and ADR can also be attributed to the annual Oktoberfest event occurring in September~\citep{herrmann2014hotel}.

\begin{table*}[htb]
    \centering
    \caption[Gini coefficients illustrating the seasonality indicators in selected cities]{Gini coefficients illustrating the seasonality indicators in selected cities. \rebuttal{The} bold notations highlight the maximum Gini indices for ADR in the city, while the italicized and underlined ones denote the minimum}
    \label{table:gini-coefficients-combined}
    \resizebox{\textwidth}{!}{
    \begin{tabular}{lcccccccccccccc}
        \toprule
        \multirow{2}{*}{\textbf{City}} & \multirow{2}{*}{\boldmath{$G_{AVC}$}} & \multicolumn{12}{c}{\boldmath{$G_{ADR}$}} \\ 
        &  & Jan & Feb & Mar & Apr & May & Jun & July & Aug & Sep & Oct & Nov & Dec \\
        \midrule
        Amsterdam & 0.146 
            & 0.033 & \minHighlight{0.010}  & 0.019 & 0.017 & 0.014 & 0.011 & 0.044 & 0.014 & \maxHighlight{0.063} & 0.060 & 0.026 & 0.047 \\
       Barcelona & 0.115  
            & 0.040 & \maxHighlight{0.108} & 0.051 & 0.027 & 0.026 & 0.026 & \minHighlight{0.008} & 0.025 & 0.029 & 0.025 & 0.026 & 0.097 \\
        Berlin &  0.158
            & 0.028 & 0.019 & 0.020 & 0.015 & 0.026 & 0.051 & 0.046 & 0.026 & 0.040 & 0.022 & \minHighlight{0.018} & \maxHighlight{0.056} \\
       Brussels & 0.116
            & 0.023 & 0.015 & 0.020 & 0.025 & 0.008 & 0.009 & \minHighlight{0.006} & 0.008 & \maxHighlight{0.081} & 0.046 & 0.028 & 0.064 \\
        Madrid &  0.079
            & 0.037 & 0.024 & 0.041 & 0.028 & 0.045 & 0.028 & 0.025 & \minHighlight{0.020} & 0.059 & 0.055 & 0.037 & \maxHighlight{0.073} \\
       Munich & 0.188
            & 0.013 & \minHighlight{0.008} & 0.029 & 0.010 & 0.020 & 0.036 & 0.033 & 0.032 & \maxHighlight{0.138} & 0.049 & 0.009 & 0.033  \\
       Paris & 0.100
            & 0.017 & \minHighlight{0.007} & 0.012 & 0.008 & 0.013 & 0.009 & 0.139 & \maxHighlight{0.150} & 0.078 & 0.039 & 0.013 & 0.044 \\  
       Vienna & 0.184
            & 0.058 & 0.016 & 0.032 & 0.020 & 0.024 & 0.021 & \minHighlight{0.015} & \maxHighlight{0.101} & 0.062 & 0.031 & 0.027 & 0.101 \\
        \bottomrule   
      
    \end{tabular}}
\end{table*}

\subsection{Summary} \label{section: seasonality summary}

\rebuttal{We examine tourism seasonality and its influence on tourist load distribution across destinations by proposing a monthly \textit{seasonality index}.
We aim to mitigate overcrowding by promoting travel during months with lower demand. We focus solely on demand estimation to gauge monthly city crowdedness.}

\rebuttal{According to research, the most effective way to measure tourism seasonality is by computing the Gini coefficients of various financial and non-financial indicators~\citep{seasonality2020croatia}. We collected data from TourMIS and Airbnb to analyze non-financial indicators such as monthly arrival visitor counts (AVC) and financial indicators like average daily rates (ADR), respectively. The seasonal demand index is a weighted combination of Gini coefficients calculated from non-financial and financial indicators.}

\rebuttal{
    We leverage data from two distinct sources to minimize any discrepancies between them. Our approach considers short-term fluctuations captured by ADR and a broader perspective offered by AVC. To maintain consistency in the overall data pattern, we normalized them separately for each data source. Traditional seasonal factors such as climate, economic development, and institutional causes such as city events were deliberately excluded from our index calculation, as they do not directly contribute to sustainable recommendations. Our findings reveal that tourism demand follows distinct seasonal trends, with peak activity typically occurring during summer.}

\rebuttal{Overall, the \textit{seasonality index} can help understand the combined impact of tourist influx and accommodation prices on the overall attractiveness of a destination throughout the year at a monthly granularity. 
When choosing destinations for specific months, the \textit{seasonality index} offers a more nuanced perspective than visitor numbers or average prices.
This can be particularly useful in identifying less crowded destinations during peak travel seasons.
}

\section{User Perception of Sustainable City Trips} \label{section: user_study}

\rebuttal{To explore how users perceive sustainability when looking for city trip recommendations, we conducted a user study involving participants with diverse travel experiences and preferences. 
This method provides valuable insights into decision-making intricacies, as established by prior research~\citep{wilson1981user}.}
By directly engaging real-world participants in simulated scenarios, we aimed to discern how tourists assess different aspects of a city trip, negotiate trade-offs, align preferences with sustainable tourism practices, and assign weights to criteria in their decision-making.

\subsection{User Study Design} \label{subsection: user study design}
The primary objective of our user study was to gain a deeper understanding of the factors influencing individuals' decisions when selecting a city for vacation and their receptiveness to sustainable recommendations for tourism destinations.
Respondents were prompted to imagine planning their next vacation to another European city and identify the most crucial factors influencing their choice of destination. 
Only a limited set of personal demographic questions pertaining to the users' age, gender, and nationality were asked to preserve the participants' privacy.

The questionnaire was designed using Qualtrics Experience Management Software~\footnote{https://www.qualtrics.com}, an online survey platform.
We recruited participants through the online crowdsourcing platform Prolific~\footnote{https://www.prolific.com}, renowned for its efficacy in subject recruitment for the scientific community~\citep{palan2018prolific}.
\rebuttal{With a focus on European participants who listed travel as one of their hobbies, the questionnaire, designed in English, was distributed to individuals through Prolific's advanced pre-screening options, and 200 final responses were collected. }
To ensure gender diversity, the preset distribution aimed for an equal representation of 50\% males and 50\% females. Demographic analyses of the survey data indicated that 33.8\% of the participants fell within the 25-34 age group, followed by 24.3\% in the 18-24 age group, and the remaining were above the age of 35.

\subsection{Transportation Sustainability Concerns}

\begin{figure}[htbp]
    {\includegraphics[width=\textwidth]{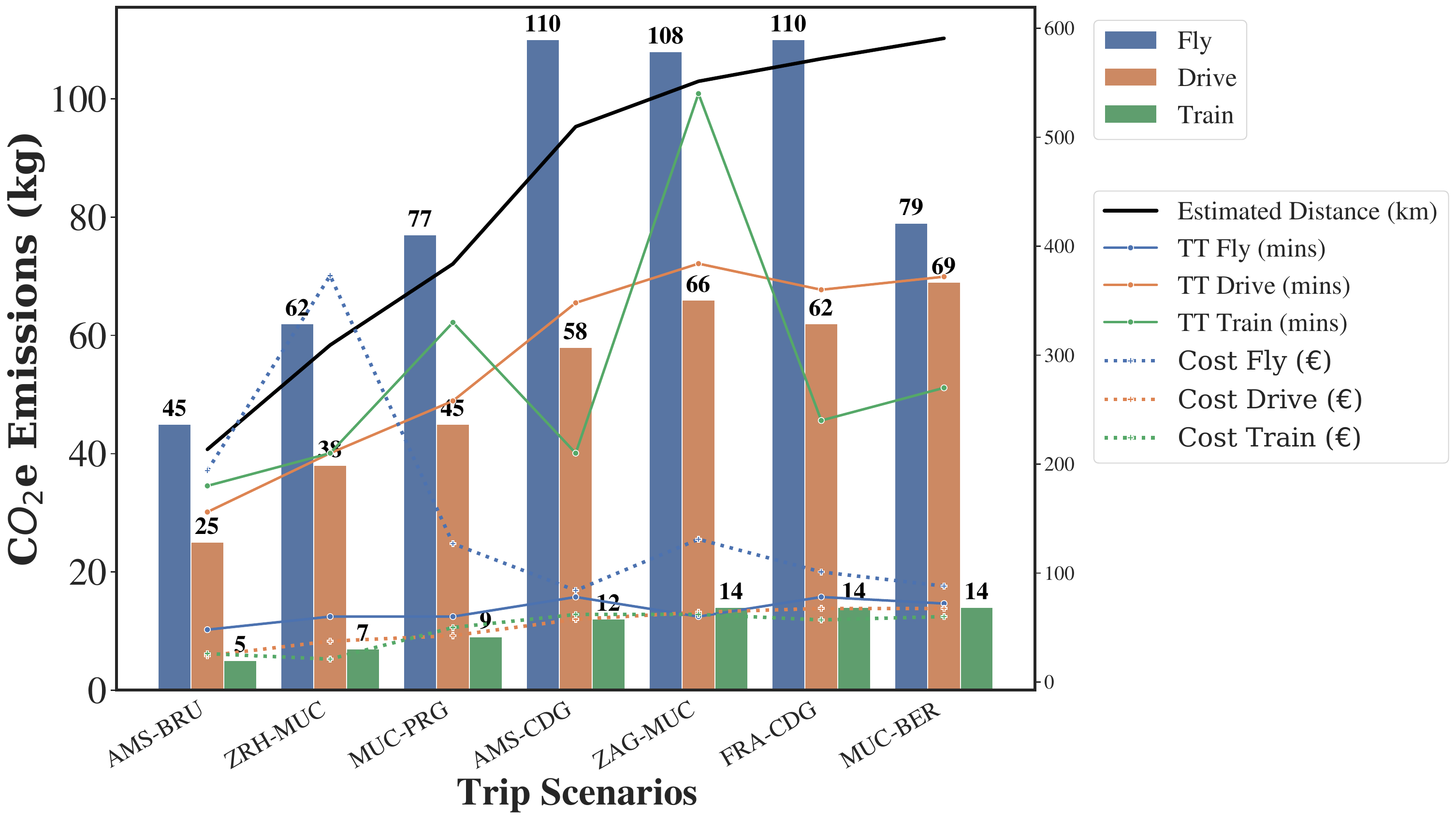}}
    \caption{\COtwoE, travel time (TT), estimated distance, and cost as shown to the users for each trip scenario, separated by transportation mode. Each city in every trip scenario is denoted by its respective IATA code}
    \label{fig: user-study-trip-scenarios}
\end{figure}

Participants were presented with seven distinct scenarios, elaborated in~\autoref{fig: user-study-trip-scenarios}. 
These scenarios illustrated trips between cities involving diverse modes of transportation --- such as train, flight, and driving with the intention of gauging their inclination towards making sustainable choices when selecting their mode of transport.
\rebuttal{The~\COtwoE~estimations and the costs tailored to each mode of transportation, were computed based on the values in~\autoref{tab:transportation-summary} and~\autoref{section: transportation tradeoff} respectively.}

\begin{figure}[htbp]
    \centering
    \includegraphics[width=0.95\textwidth]{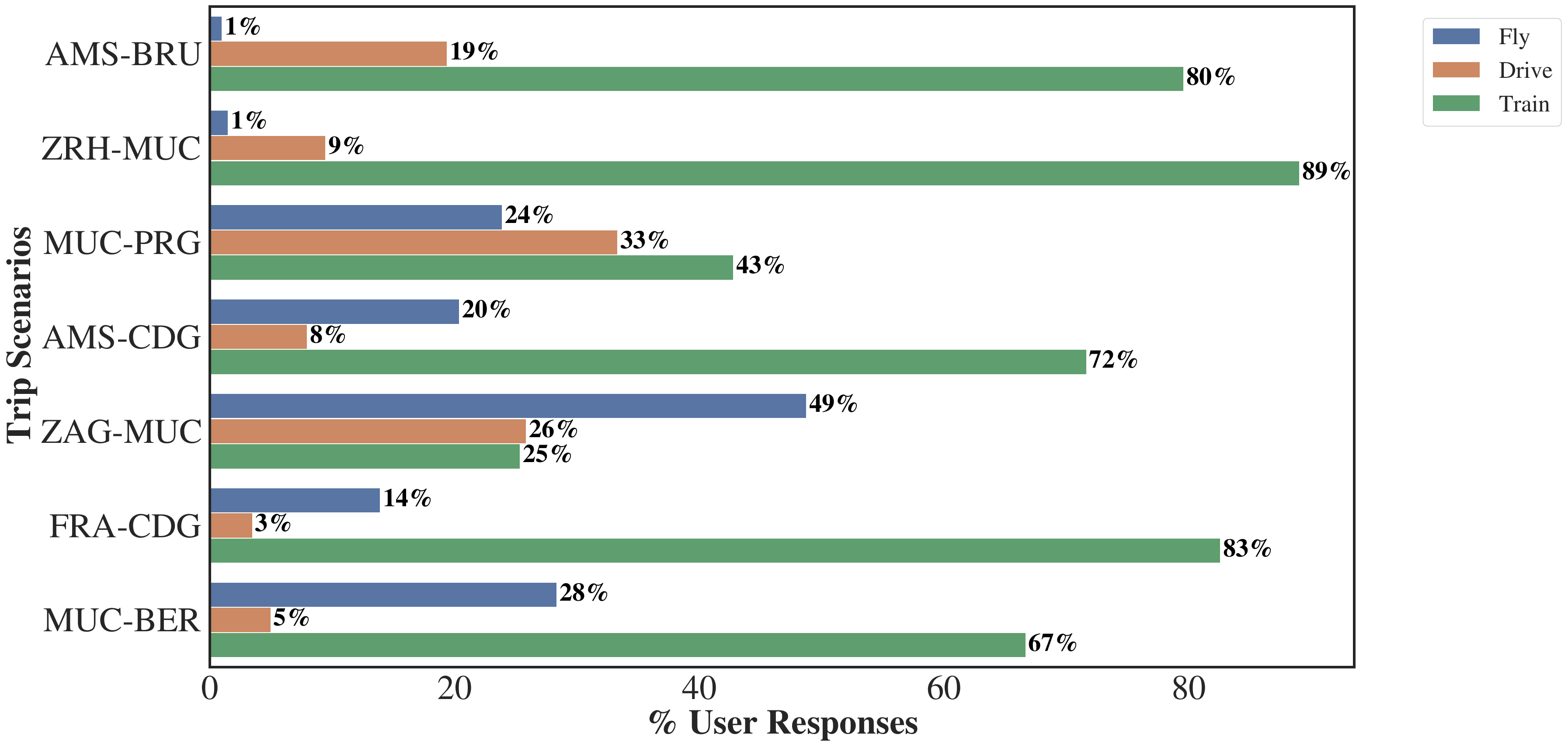}
    \caption{Distribution of user responses (\%) for different transport modes in various trip scenarios. Each city in every trip scenario is denoted by its respective IATA code}
    \label{fig:trip_scenarios_results}
\end{figure}

~\autoref{fig:trip_scenarios_results} provides a summary of the mode distribution for various city pairs, indicating the percentages of user responses for different transportation modes (train, drive, and fly). 
Key takeaways include preferences for specific modes for each trip, reflecting the distribution of travel choices considering the associated distances. Notably, train travel dominates in several instances, with variations depending on the city pairs and their distances. 

\begin{figure}[htbp]
    \centering
    \includegraphics[width=0.95\textwidth]{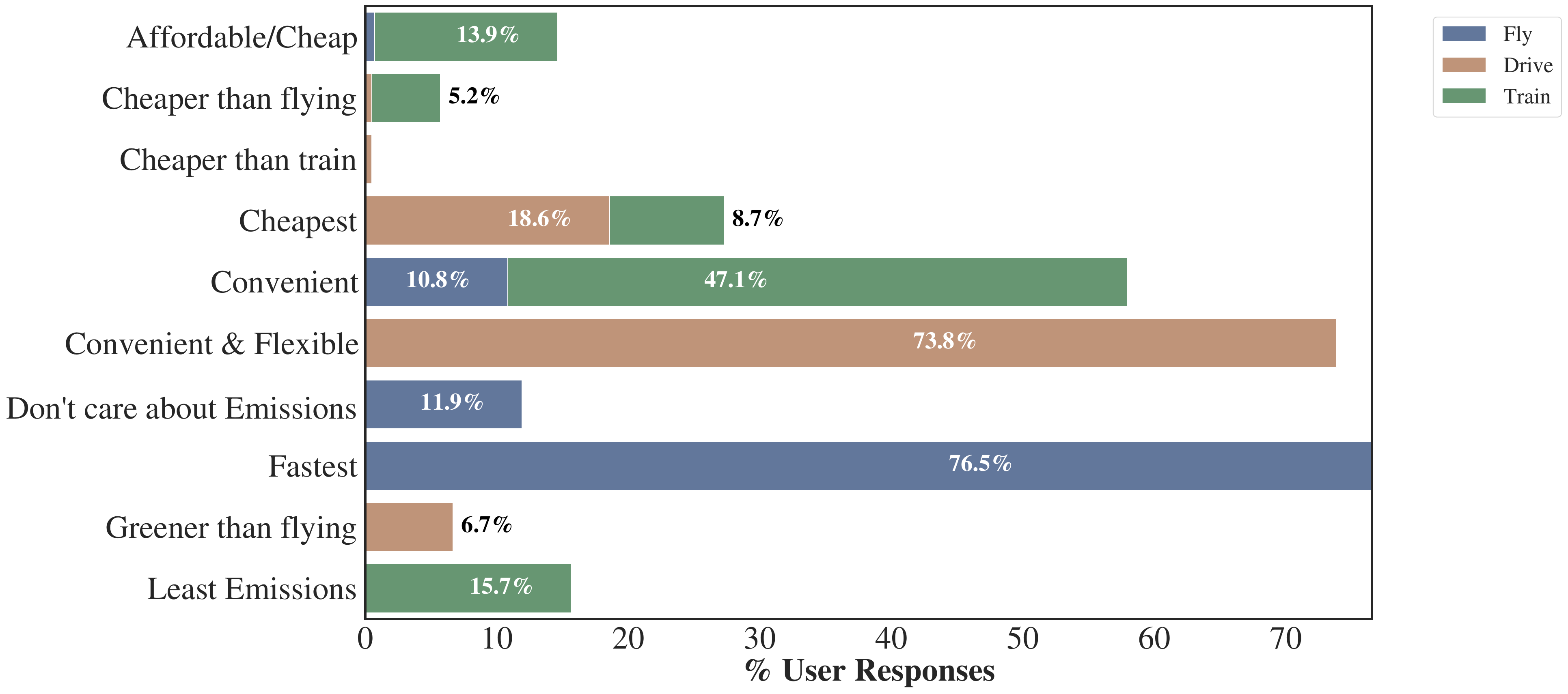}
    \caption{Reasonings provided by respondents when choosing modes of transport across all trips}
    \label{fig:trip_tradeoff_reasonings_heatmap}
\end{figure}

We investigated the reasons behind selecting trade-offs for various transportation modes in each trip scenario, as illustrated in~\autoref{fig:trip_tradeoff_reasonings_heatmap}. 
Our analysis reveals valuable insights from survey responses. For train travel, a significant 47.1\% of respondents prioritize convenience, while 15.65\% emphasize selecting the mode with the least~\tunar{\COtwoE}~emissions. Affordability is a critical factor for 13.91\% of participants, and 8.70\% opt for the cheapest option available. Additionally, 5.22\% favor trains over flights for cost considerations. 
In contrast, driving is primarily chosen for its convenience and flexibility, with an overwhelming 73.81\% of respondents highlighting this aspect. Affordability remains a factor, as 18.57\% opt for the cheapest driving option. Some respondents (6.67\%) perceive driving as environmentally better than flying. 
\rebuttal{Flying is chosen by 76.53\% for its speed, while 11.91\% remains unconcerned about \tunar{\COtwoE}~emissions.} 
The findings underline the multifaceted nature of decision-making, encompassing convenience, environmental concerns, and cost considerations across different transportation modes.
\rebuttal{Our results are consistent with those presented by~\citet{avogadro2021replacing}, which also indicates a preference for public transport over flying for shorter distances.}

\subsection{Understanding Trade-offs}
The user study also aimed to explore the various trade-offs associated with trip planning.
Participants were presented with Likert scale~\citep{joshi2015likert} statements from \textit{"not at all important"} to \textit{"extremely important"} to gauge their agreement levels to the following statements:

\begin{enumerate}[label=\subscript{S}{{\arabic*}}:]
    \item Importance of presence of off-season discounts.
    \item Climate at the destination.
    \item Cost savings by traveling during the off-season.
    \item Visiting the city during its best travel time, even during the peak tourist season.
    \item Overall attractiveness of the destination.
    \item The destination in terms of unique attractions, points of interest, etc., even if that means they are very popular.
    \item Cities that are widely popular, even if they might be crowded.
\end{enumerate}

\begin{figure}[htbp]
    {\includegraphics[width=\textwidth]{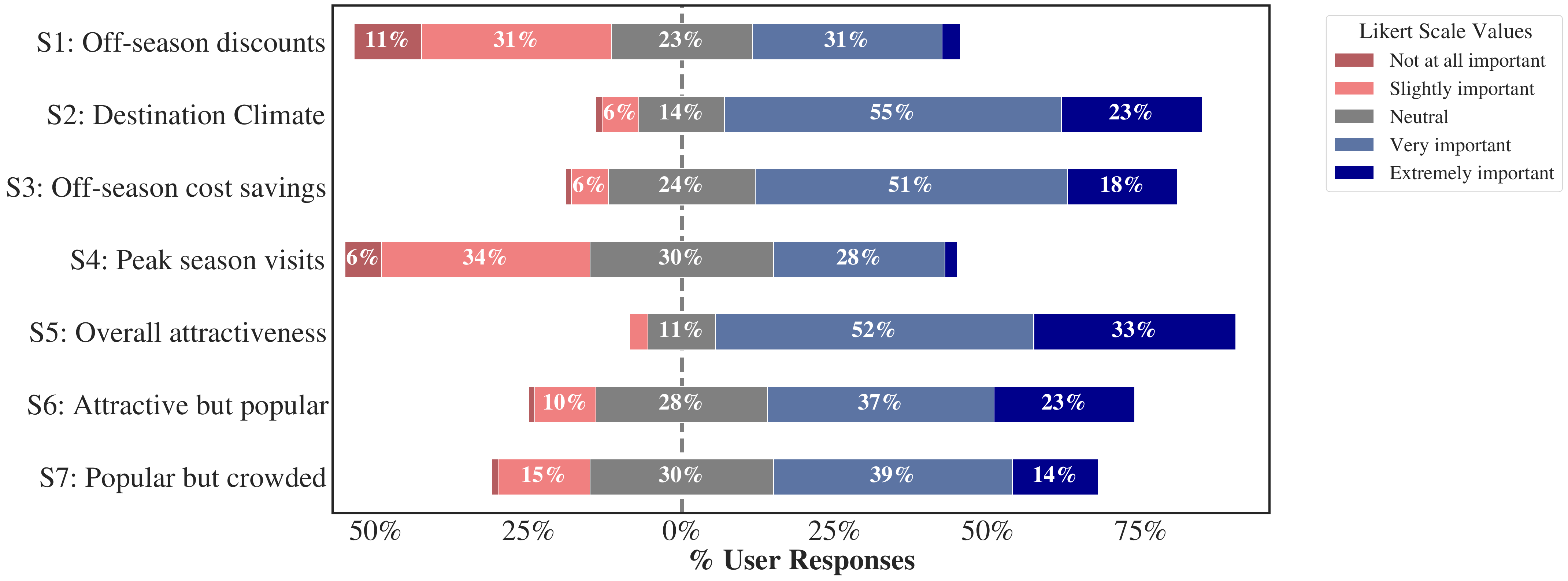}}
    \caption{\rebuttal{Degree of the importance of the various factors influencing decision-making during trip planning}}
    \label{fig:other_trade_offs_results_viz}
\end{figure}

The analysis of user responses, as depicted in~\autoref{fig:other_trade_offs_results_viz}, provides insights into the trade-offs users are willing to make when considering various attributes in travel decision-making. Notably, for the attribute \textit{"popular but crowded,"} a minimal percentage (2.99\%) strongly agreed, while the majority (31.34\%) agreed. Conversely, for the attribute \textit{"attractive but popular"}, a significant proportion (55.22\%) agreed, indicating a higher tolerance for popularity in the pursuit of attractiveness. The consideration of \textit{"overall attractiveness"} saw 51.24\% in agreement. Respondents expressed varying opinions on the \textit{"visiting in peak season"}, with 1.99\% strongly disagreeing and 33.83\% expressing disagreement. Furthermore, factors such as \textit{"off-season cost savings"}, \textit{"Climate at the destination"} during the time of travel, and \textit{"off-season discounts"} revealed nuanced preferences, with notable percentages in agreement (51.74\%, 37.31\%, 39.30\%, respectively) and distinct proportions holding dissenting views. These results contribute valuable insights into the varied considerations influencing users' travel preferences.

\section{Societal Fairness} \label{section: s-fairness}

\rebuttal{This section explores the concept of \tunar{Societal Fairness (\SF)}, which is a crucial component in evaluating city trip recommendations. \tunar{\SF} seeks to ensure that tourism benefits and impacts are distributed fairly among a broad range of stakeholders, including not only tourists, service providers, and platforms, but also non-participating entities such as residents, locals, and the environment~\citep{ashmi_umap_dc_2023,banik2023fairness}.}

\rebuttal{
To evaluate \tunar{\SF} across different destinations, our methodology combines insights from established data sources with findings from a comprehensive user study, providing a comprehensive perspective.
We define the \tunar{\textit{Societal Fairness Indicator (\SFI)}} by quantifying the overall impact of a destination on both the environment and society relative to the user's starting location.}

\subsection{Defining the S-Fairness Indicator}

Each destination $\Cityi$ during the month $j$ accessible from the user's origin is assigned an \tunar{\SFI}, denoted as $\SFindexCij$. This indicator is determined through a weighted combination of three essential components ---
\begin{enumerate}
    \item \textit{Emissions trade-off index} $\EmissionScorei$ from~\autoref{section: transportation} with its normalized weight represented by $\alpha$
    \item \textit{Popularity index} $\popScoreCi$ from~\autoref{section: popularity} with its normalized weight denoted as $\beta$
    \item \textit{Seasonality index} $\sigmaCij$ from~\autoref{section: seasonality} with its normalized weight represented by $\Gamma$.
\end{enumerate}

The formulation is expressed as:

\begin{equation}
\label{eq: S-Fairness Indicator}
\SFindexCij = \alpha \cdot Z(c_{i}) + \beta \cdot \rho(c_{i}) + \Gamma \cdot \sigma(c_{i}^{j})
\end{equation}

Here, $\SFindexCij$ falls within the range of zero to one, where a higher indicator signifies a more adverse impact on society. The allocation of weights to these indicators reflects the significance of considering emissions, popularity, and seasonality in evaluating the societal fairness of a destination. This approach, integrating both quantitative and user-centered perspectives, strengthens the effectiveness of our \tunar{\SFI}.

\subsection{Learning Weights of the Indices} \label{section: learning weights}

\rebuttal{One of the main goals of the user study described in~\autoref{subsection: user study design} was to understand the importance of different factors that the users consider when choosing travel destinations. 
Participants were requested to prioritize their transportation choices based on travel time, \tunar{\COtwoE} footprint, and cost.
They also provided feedback on other essential factors, such as the number of points of interest, Google image search values, and the total number of Tripadvisor reviews and opinions. Additionally, they indicated the importance of costs associated with available accommodation and crowd levels in the city during their visit. 
These factors align with the coefficients of the \textit{"emissions trade-off index"} $\EmissionScorei$ in~\autoref{eq: emission score}, \textit{"popularity index"} $\popScoreCi$ in~\autoref{eq: popularity score}, and  \textit{"seasonality index"} $\sigmaCij$ in~\autoref{eq: seasonality index} respectively.}

Finally, the participants were prompted to assess the influence of various factors on their decision-making process to estimate the weights associated with the combined \tunar{\SFI} $\SFindexCij$, as illustrated in~\autoref{eq: S-Fairness Indicator}. They were explicitly asked about the impact of factors such as having a lower~\tunar{\COtwoE}~footprint, opting for a less famous city, and avoiding the busiest time of the year when making travel decisions.
All the responses were gathered using a 5-point Likert scale.

\begin{figure}[h]
    \centering
    {\includegraphics[width=0.8\textwidth]{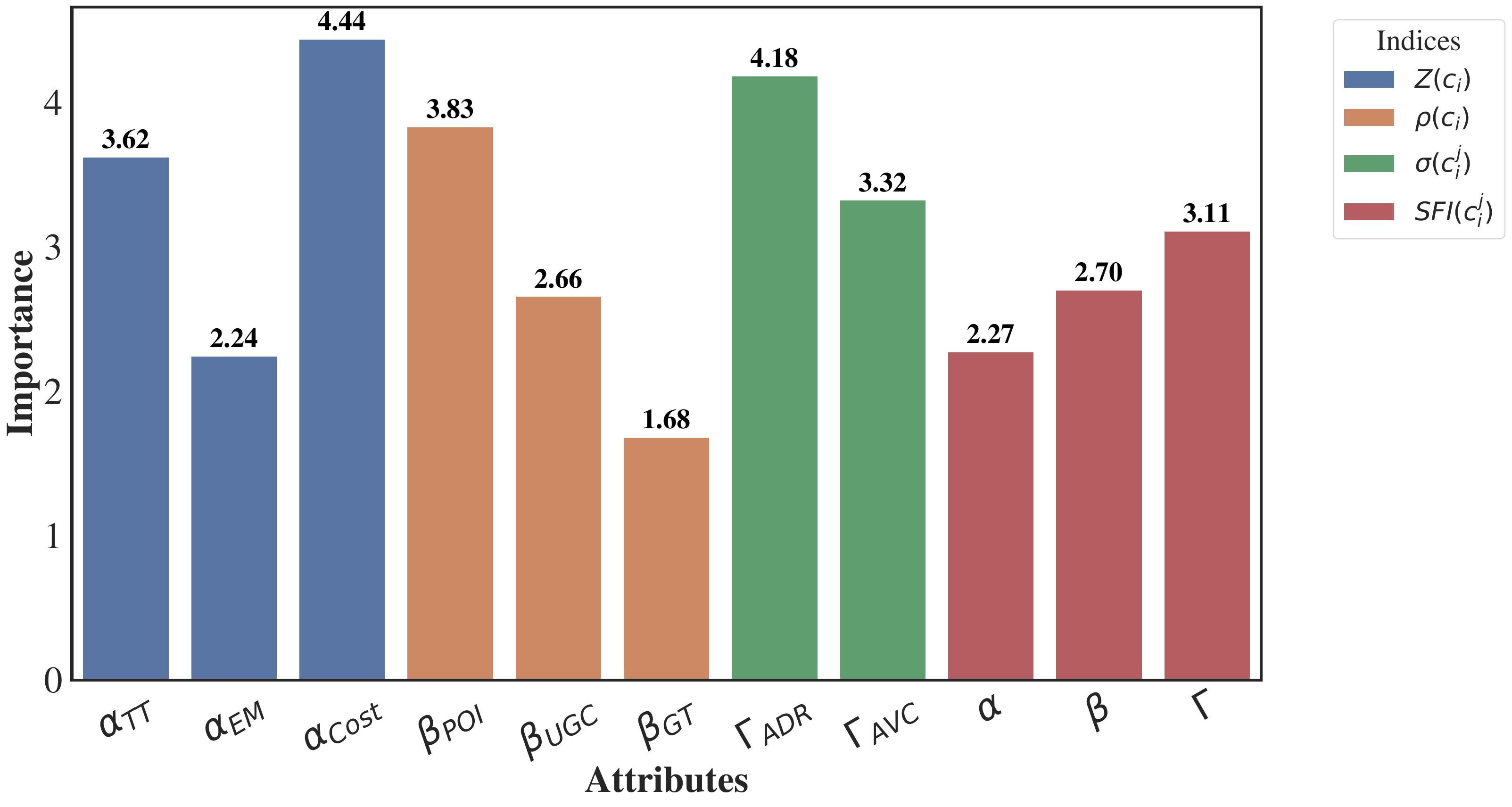}}
    \caption{Distribution of the absolute values of weights, obtained through the weighted average of Likert scale results}
    \label{fig:user-study-weights}
\end{figure}

We calculated weighted averages on Likert scales, spanning from  \textit{"not important at all"}, having a minimum weight of 1, to \textit{"extremely important"}, with a maximum weight of 5, enabling us to identify patterns in the composite indicator. The distribution of the absolute values of these weights, obtained through the weighted average of Likert scale results, are shown in ~\autoref{fig:user-study-weights}.
Additionally, we applied min-max normalization~\citep{patro2015normalization} to normalize these averages within each category, enabling us to gauge their relative significance in their respective categories. 
These normalized weights offer valuable insights into participants' preferences and priorities, shedding light on the factors that significantly influence their decision-making when choosing travel destinations.
After incorporating the normalized weights, the expressions for the \textit{emissions trade-off index}, \textit{popularity index}, and \textit{seasonality index} can be revised as follows:

\begin{equation}
    \label{eq: emission score weighted}
    \EmissionScorei = 0.352 \cdot \tau_{TT}({\Cityi}) + 0.218 \cdot \tau_{EM}({\Cityi}) + 0.431 \cdot \tau_{Cost}({\Cityi})
\end{equation}
\begin{equation}
    \label{eq: popularity score weighted}
    \rho(c_{i}) = 0.469 \cdot \pi_{POI} + 0.325 \cdot \pi_{UGC} + 0.206 \cdot \pi_{GT}
\end{equation}
\begin{equation}
    \label{eq: seasonality score weighted}
    \sigmaCij = 0.443 \cdot G_{AVC}^{(i)} + 0.557 \cdot G_{ADR}^{(i)(j)} 
\end{equation}

Combining the weighted formulations presented in \autoref{eq: emission score weighted}, \autoref{eq: popularity score weighted}, and \autoref{eq: seasonality score weighted}, we obtain the updated values for the \tunar{\SFI} $\SFindexCij$ as follows:
\begin{equation}
    \label{eq: S-fairness index weighted}
    \SFindexCij = 0.281 \cdot \EmissionScorei  + 0.334 \cdot \popScoreCi +  0.385 \cdot \sigmaCij
\end{equation}

The weights here represent the importance assigned to different attributes within each category. Notably, in the emission trade-off category $\EmissionScorei$, the highest weight is given to the cost attribute ($\alphaCost$), indicating that users prioritize the cost factor when evaluating emission indices. In the popularity category $\popScoreCi$, the points of interest ($\betaPOI$) attribute carries the highest weight, suggesting that users prioritize locations with significant points of interest. However, in the \tunar{\SFI} category $\SFindexCij$, the weight distribution is relatively balanced among the attributes $\alpha$, $\beta$, and $\Gamma$, indicating a more equitable consideration of these factors.

The composite \tunar{\SFI} $\SFindexCij$ assigned to each city $\Cityi$ for the month $j$ signifies the extent of negative environmental impact associated with traveling to the city from users' starting points. A lower value of $\SFindexCij$ indicates a more environmentally friendly choice and lesser harm caused. Our objective is to encourage individuals to visit cities with lower \tunar{\SFI} relative to their starting points, aiming to minimize the adverse effects of tourism on the environment and promote sustainable and responsible tourism practices.

\subsection{Integrating S-Fairness Indicator into TRS}\label{section: validation}

\begin{figure*}[htbp]
    \centering
    {\includegraphics[width=\textwidth]{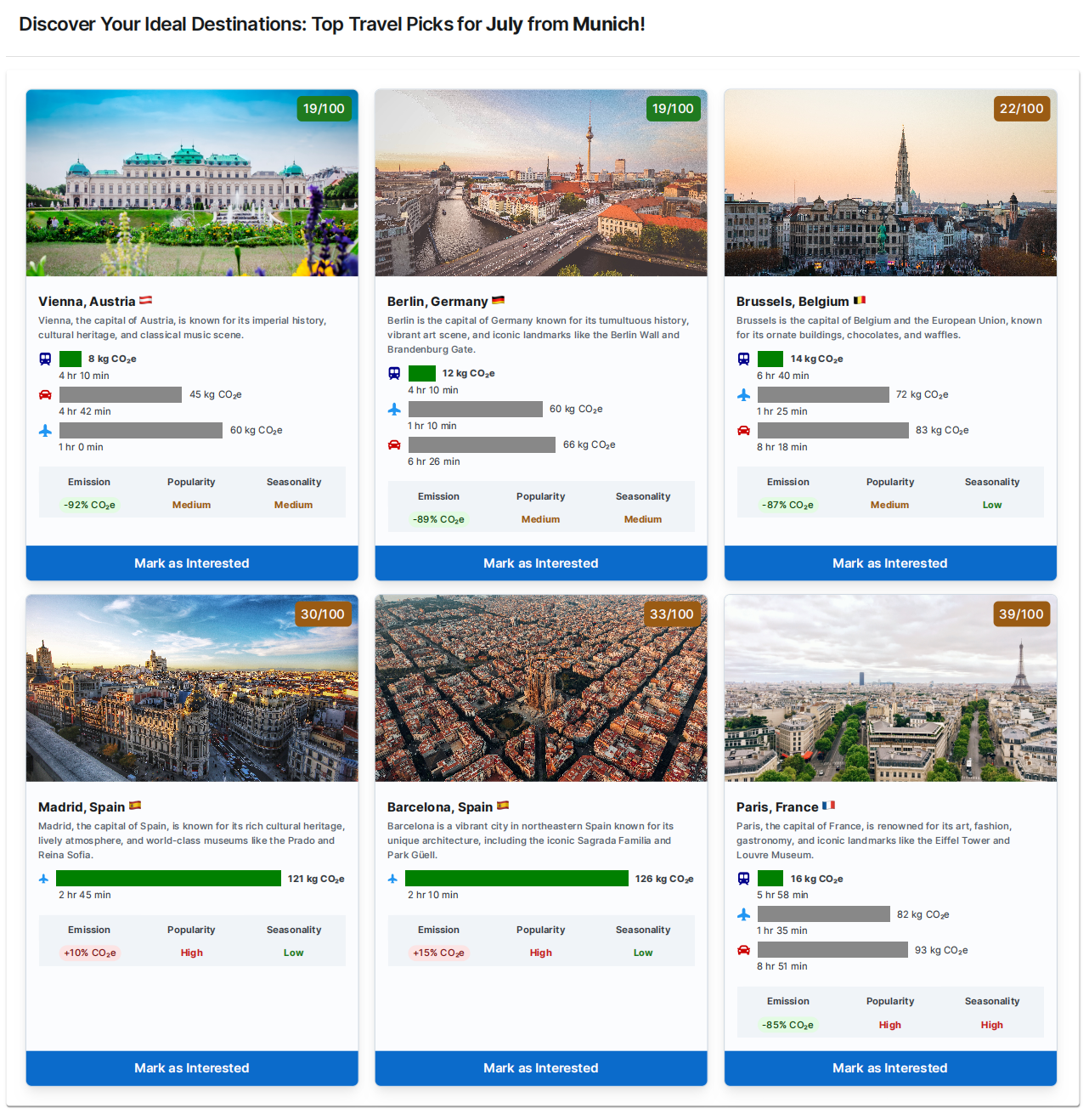}}
    \caption{
        \rebuttal{Snapshot of the sample user interface showing top travel recommendations from Munich for July while highlighting the \tunar{\SFI} in the top right corner, shown to the participants during the user study}}
    \label{fig: ui}
\end{figure*}

\rebuttal{In this section, we present a separate user study demonstrating how our proposed \tunar{\SFI} can be integrated into a TRS in real-world scenarios.
To evaluate the effectiveness of our \tunar{\SFI}, we recruited 200 European residents through Prolific, ensuring gender balance and an interest in travel.
Participants were shown a user interface snapshot~\autoref{fig: ui} that displayed the top travel destinations from Munich for July. It included a photograph and brief overview of each destination, as well as travel time and \tunar{\COtwoE} emissions information for different modes of transportation from Munich. 
To aid comprehension, each city was labeled with popularity and seasonality tags denoting their respective levels of popularity and monthly seasonality.
Cities within the top 5 percentile of their respective popularity and seasonality indices were categorized as \textit{high}, those in the top 50 percentile as \textit{medium}, and the rest as \textit{low}. 
We assigned an overall \tunar{\SFI} out of 100 to each city, displayed in the top right corner. This was derived by multiplying our \tunar{\SFI} from~\autoref{section: learning weights} by 100. It represents the city's overall sustainability status when traveling from Munich.}

Following the presentation, participants were tasked with expressing their opinions on the provided statements using a 5-point Likert scale~\citep{joshi2015likert} ranging from \textit{"strongly disagree"} to \textit{"strongly agree"}. The statements covered the following aspects:
\begin{enumerate}[label=\subscript{S}{{\arabic*}}:]
    \item The assigned \tunar{\SFI} scores accurately reflect the sustainability of the showcased city destinations.
    \item Cities with lower \tunar{\SFI} scores are perceived as more appealing for travel.
    \item \tunar{\SFI} are deemed helpful in facilitating informed decisions about preferred travel destinations.
\end{enumerate}

\begin{figure}[htbp]
    {\includegraphics[width=\textwidth]{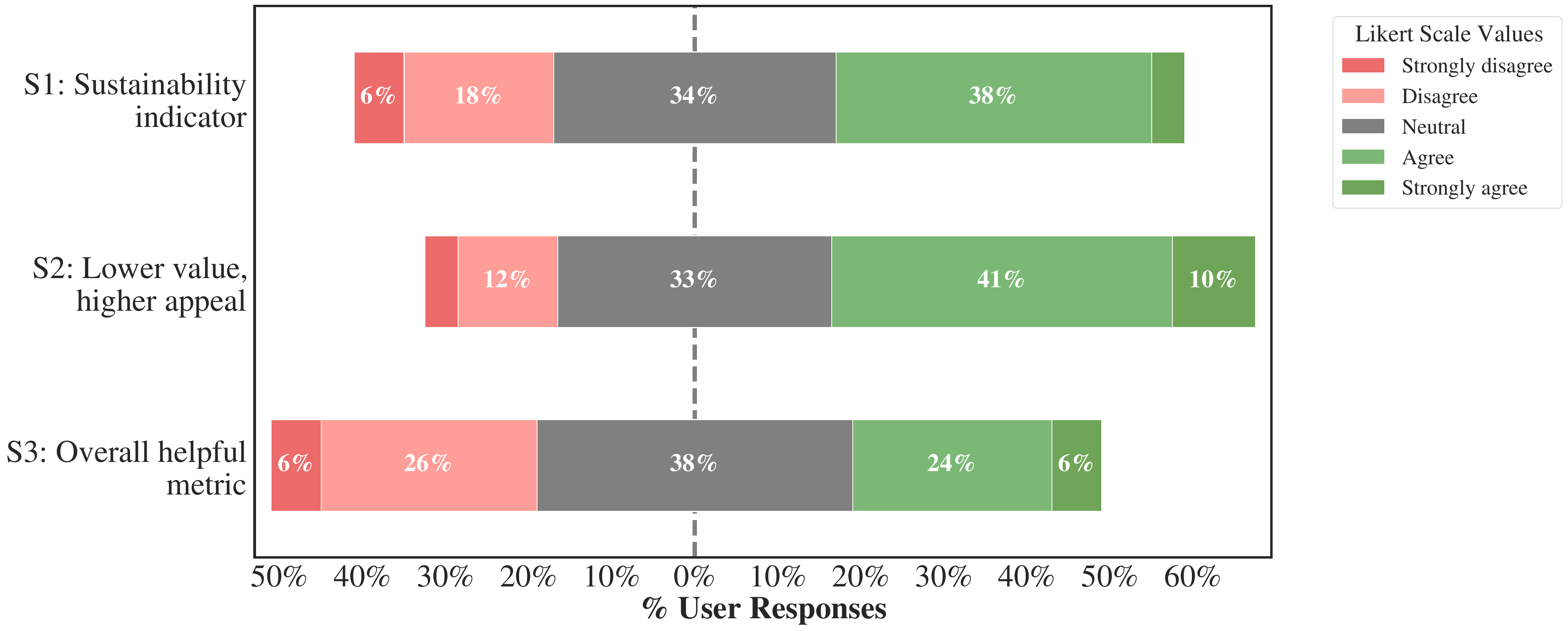}}
    \caption{Results of the user study to validate the usefulness of the \tunar{\SFI} (SFI), showing a mostly positive response}
    \label{fig: sf_val_results_likert}
\end{figure}

The results are depicted in~\autoref{fig: sf_val_results_likert}, revealing a generally positive reception to the \tunar{\SFI} as a \textit{sustainability indicator}. A majority of users (72\%) expressed neutrality or agreement, with 4\% of the users expressing very strong agreement. 
The \textit{"lower value, higher appeal"} metric showed a similar trend, with 84\% ranging from neutral to solid agreement, including strong agreement, suggesting that a lower value of \tunar{\SFI} correlates with higher user appeal. 
\rebuttal{However, the \textit{"overall helpful metric"} elicited more varied responses, with 30\% of users agreeing and 32\% disagreeing. Given the marginal difference between agreement and disagreement, particularly concerning this metric, it is possible that users did not fully grasp the concept presented in the interface. Improving the representation or providing clearer motivation in the user interface may address this issue in future iterations. This limitation of our study will be addressed in subsequent research efforts.
}
Overall, there is a trend of approval across all metrics, with even the least favorable response showing a majority of users being neutral to strongly agreeing on the value of the metrics. 

\rebuttal{
While~\autoref{fig: ui} serves as an illustrative user interface, the potential of this work extends to the development of a practical application. Such an application could present destinations and their \tunar{\SFI}, incorporating up-to-date information like real-time connectivity and cost considerations. Moreover, it could integrate additional metrics such as accommodation availability and environmental factors like climate and air quality indices. This integration, coupled with user preferences, would enable the provision of real-time recommendations that align with sustainability principles and individual user preferences.}

\rebuttal{Additionally, as depicted in~\autoref{fig: ui}, destinations are currently ordered by the \tunar{\SFI}. However, in practical scenarios, users may prefer sorting destinations by emissions, popularity, and seasonality indices, given their varying importance to different individuals.
Introducing a user interface that allows sorting and filtering by the individual components while making the associated trade-offs explicit would enhance the flexibility of the application.
Similarly, the current representation of popularity and seasonality labels (e.g., \tunar{\textit{low}}, \tunar{\textit{medium}}, or \tunar{\textit{high}}) in~\autoref{fig: ui} could be refined through discrete categories to offer meaningful and actionable insights, contributing to a more user-friendly and informative interface.
We intend to dive deeper into communicating sustainable recommendations to users as part of our future research.}

\section{Conclusion} \label{section: conclusion}

\rebuttal{In conclusion, our research introduces a novel approach for assigning a sustainability indicator \tunar{(\SFI)} for city trips accessible from the users' starting point, integrating \tunar{\COtwoE}~emission analysis, destination popularity, and seasonal demand to provide well-rounded and sustainable city trip suggestions.
The theoretical implication of this concept lies in extending sustainability beyond environmental concerns to ensure equitable benefits distribution among stakeholders.
Our methodology, validated through a user study, showcases the practicality and effectiveness of the model in providing well-rounded and sustainable city trip suggestions.
The findings indicate that while there is a general awareness of sustainability, tourists often prioritize convenience and personal preferences over sustainable choices. This gap highlights the need for more effective communication and education strategies to promote \tunar{\SF} in city trip planning.}

\rebuttal{Our study is particularly interesting for stakeholders such as travelers seeking sustainable travel options, tourism industry professionals looking to promote responsible tourism practices, and policymakers aiming to implement sustainability initiatives in urban tourism. 
While the implementation of our system is feasible leveraging existing data sources and technologies, challenges such as data availability and user adoption may need to be addressed.
The absence of personalization features in the current version presents an opportunity for further research to explore this area to better understand user preferences and their decision-making processes.
Furthermore, our sustainability metric can also be extended to include other impacting factors such as accommodation availability, or environmental factors such as climate and air quality index.}
\rebuttal{
In summary, our study integrates sustainability and societal fairness into TRS, laying the foundation for a system that aligns eco-friendly travel recommendations with the evolving traveler needs and preferences.}

\bibliography{main.bib}%

\end{document}